\documentclass[preprint,prd,aps,nofootinbib]{revtex4}
\usepackage{amsmath}
\usepackage{graphicx}
\usepackage{tikz}
\usepackage{diagbox}
\usepackage{slashed}
\usepackage{float}
\usepackage{color}
\begin{document}
\title{{Semirelativistic} study on the semileptonic decays of $B_q$ mesons to orbital excited heavy Tensors}
\author{Wen-Yuan Ke$^{1,2,3}$\footnote{20218015001@stumail.hbu.edu.cn},
Su-Yan Pei$^{1,2,3}$, Tianhong Wang$^4$,
Guo-Li Wang$^{1,2,3}$\footnote{wgl@hbu.edu.cn, corresponding author}}
\affiliation{$^1$ Department of Physics, Hebei University, Baoding, 071002, China\\
$^2$  Hebei Key Laboratory of High-Precision Computation and Application of Quantum Field Theory, Baoding 071002, China\\
$^3$ Hebei Research Center of the Basic Discipline for Computational Physics, Baoding 071002, China\\
$^4$ School of Physics, Harbin Institute of Technology, Harbin 150001, China}
\begin{abstract}
Based on the method of solving the complete Salpeter equation, we study the semileptonic decays of a  $0^-$ heavy meson to $1P$, $2P$, or $3P$ heavy tensor mesons, $B_q \to (\bar c q)(nP) \ell^+ \nu_\ell$ $(q=u,d,s,c;n=1,2,3)$. The obtained branching ratio of $\mathcal{B} (B \rightarrow  D_2^{\star}(2460)\ell^{+} \nu_{\ell})$ agrees with the experimental data. We predict $\mathcal{B}\left(B_s^{0} \rightarrow D_{s2}^{\star-}(1P) \ell^{+} \nu_{\ell}\right)$$=$$3.76\times 10^{-3}$ and $\mathcal{B}\left(B_c^+ \rightarrow \chi_{c2}(1P)\ell^{+} \nu_{\ell}\right)$$=$$1.82\times 10^{-3}$. The branching ratios of decays to $2P$ and $3P$ final states are found to be very small. The ratios $\mathcal{R}(\bar{D}_{2}^{\star 0})=0.045$, $\mathcal{R}({D}_{s2}^{\star})=0.048$ and $\mathcal{R}({\chi}_{c2})=0.059$ are also obtained.
This study focuses on the contribution of relativistic corrections.
The wave function of the pseudoscalar includes non-relativistic $S$-wave and relativistic $P$-wave. While for a tensor, it contains non-relativistic $P$-wave and relativistic {$P$}, $D$ and $F$ waves in its wave function.
We find the individual contributions of relativistic partial waves are significant in the decay $B \to D_2^{\star }(2460)\ell^{+} \nu_{\ell}$, but the overall contribution of the relativistic effect is $24.4\%$, which is small due to cancellation. Similarly, for the decay $B_s^{0} \rightarrow D_{s2}^{\star-}(1P) \ell^{+} \nu_{\ell}$, the contribution of the relativistic effect is $28.8\%$. While for $B_c^+ \rightarrow \chi_{c2}(1P)\ell^{+} \nu_{\ell}$, the individual contributions of relativistic partial waves and the overall relativistic correction are both small, the later of which is $22.1\%$.
\end{abstract}
%\ pacs{12.60.Jv, 14.80.Cp}
%\ keywords{BS equation, Bottom meson}
\maketitle
\section{Introduction}
In the past few years, the semileptonic decays of bottom mesons induced by $b\to c$ have attracted a lot of research interest both in theory \cite{Kang:2018jzg,Dingfelder:2016twb,Aliev:2006gk,lv2,lv3} and in experiment \cite{BaBar:2008dar,Belle:2021idw,BaBar:2011sxq,D0:2007ukf,LHCb:2020ayi,LHCb:2021tdf}, since such decays are important for the studies of the Cabibbo-Kobayashi-Maskawa (CKM) matrix element $V_{cb}$ \cite{Belle-II:2023jgq,bibi2}, CP violation \cite{BaBar:2001,Belle:2001}, probing new physics \cite{Fajfer,lv1}, etc.
So far, many processes have been extensively studied, such as the semileptonic decays of $B$ to $D$ or $D^{\star}$. However our knowledge on the final state being an orbitally excited state is still insufficient. For example, there is the long-lived `$1/2$ vs $3/2$' puzzle \cite{scora,puzzle,colangelo,bigi,me05} in $B$ semileptonic decays to orbitally excited states.

Among the orbitally excited states, the $2^+$ tensor meson is a very complex one. There are significant differences between theoretical results on $B \rightarrow  D_2^{\star}(2460)\ell^{+} \nu_{\ell}$, a few results are in good agreement with experimental data, see Table \ref{All result} in this article for details. The relativistic correction of an excited state is greater than that of the ground state \cite{wangv}, so one possible reason for the inconsistency between theory and experiment is that the relativistic correction was not well considered. Therefore, in this article, we will give a {semi-relativistic study} of the semileptonic decays, $B_q \to (\bar c q)(nP) \ell^+ \nu_\ell$ $(q=u,d,s,c;n=1,2,3)$,  where $B_q$ is a pseudoscalar meson, and the final meson ($\bar{c}q$) is a tensor meson. The processes with highly excited $2P$ and $3P$ final states are also included, as we know almost nothing about them.

In this paper, we will solve the instantaneous Bethe$-$Salpeter (BS) equation \cite{BS equation}, which is also called Salpeter equation \cite{Salpeter}, to obtain the Salpeter wave functions for pseudoscalar and tensor mesons. Compared with the non-relativistic Schrodinger equation, the BS equation is a relativistic dynamic equation for bound states. As it is very complicated, we have to make approximation before solving it. Salpeter equation is its instantaneous version, and the instantaneous approximation is suitable for heavy mesons. {Due to instantaneous approximation, this method is no longer strictly relativistic, but a semi-relativistic approach.} We have solved the complete Salpeter equation without further approximations \cite{wang0-,wang2+}. Since Salpeter equation itself does not provide the form of wave functions, we give the general expression of the {Salpeter wave function} for a meson according to its $J^P$ quantum number, where the unknown radial wave functions are the solution of Salpeter equation.
The { Salpeter equations} satisfied by mesons with different $J^P$ need to be solved separately, see Ref.\cite{changwang1} for example.

It is known that some particles are not pure wave states, such as $\psi(3770)$ which is $S-D$ mixing state \cite{eichten}.
In our method, a meson wave function contains different partial waves, each of which has the same $J^P$. It is found that \cite{part wave}, in {our semi-relativistic method, or any complete-relativistic method}, similar conclusions applies to all particles, that is, all particles are not composed of pure waves, but contain other partial waves in addition to the main one.
The main wave provides the non-relativistic contribution, while others give relativistic corrections. Taking $B_c$ meson as an example, the $S$-wave is its main wave which is non-relativistic, while $P$-wave is the relativistic correction term \cite{part wave}.

Although we can calculate the ratios of different partial waves, which reflect the relativistic effect \cite{part wave}, they do not represent the size of the relativistic effect in the transition it participates in. In a transition process, it is necessary to calculate the overlapping integral of the initial and final state wave functions. In this case, the relativistic correction becomes complex and requires careful study.
The main contribution may not necessarily come from the non-relativistic partial wave, but may come from the relativistic ones.
This phenomenon motivates us to study the role of various partial waves in different decays.
Previously, we have studied the contribution of various partial waves in strong \cite{Liu:2022kvs} and electromagnetic transitions \cite{liwei,peisy}.
In this article, we will study their performance in the weak transition.

{In Sect.II, we introduce the Bethe-Salpeter equation and its instantaneous version, that is, Salpeter equation. In Sect.III, the wave functions including different partial waves of initial $0^-$  and final $2^+$ mesons are given. We also show the details to solve the Salpeter equation of $2^+$ state. In Sect.IV, taking the semileptonic decay $B^+ \to \bar D_{2}^{\star }(2460)^0 \ell^+ \nu_\ell$ as an example, we show with our method how to calculate the transition matrix element. In Sect.V, we present the ratios of different partial waves in the wave functions of $0^-$ and $2^+$ mesons, and the results of semileptonic bottom meson decays. The contributions of different partial wave and discussions are also given.}

\section{Introduction of Bethe-Salpeter equation and Salpeter equation}
{ The BS equation is Lorentz covariant within the framework of quantum field theory, which describes the relativistic two-body bound state.
The BS equation for a bound state composed of quark 1 and antiquark 2 is generally expressed as,
\begin{equation}\label{BS eq}
(\slashed p_1 - m_1)\chi_{_P}(q)(\slashed p_2 + m_2 )=i \int \frac {d^4k}{(2 \pi)^4} V(P,k,q) \chi_{_P}(k),
\end{equation}
where $p_1$ represents the quark's momentum, $p_2$ is the antiquark's momentum; $m_1$ and $m_2$ are the masses of the quark and antiquark, respectively; $\chi_{_P}(q)$ denotes the BS wave function of the meson; $V(P,k,q)$ represents the integral kernel of the BS equation; $P$ is the meson's total momentum, and $q$ is the relative momentum between the quark and antiquark. We have the following relation
\begin{align*}
p_1 & = \alpha_1 P + q, & \alpha_1 & = \frac{m_1}{m_1 + m_2}, \nonumber\\
p_2 & = \alpha_2 P - q, & \alpha_2 & = \frac{m_2}{m_1 + m_2}. \nonumber
\end{align*}

The BS equation is a four-dimensional integral equation, which is very difficult to solve. So various approximation have been developed to solve it. Among them, the instantaneous approximation is the most frequently used one, which was first proposed by Salpeter and is suitable for heavy mesons.
In the center of mass system, with the condition of instantaneous approximation, the interaction kernel is simplified as
\begin{equation}\label{instantaneous}
V(P,k,q)\sim V(\vec{k},\vec{q})=V(\vec{q}-\vec{k}),
\end{equation}
then the four-dimensional BS equation can be reduced to the three-dimensional Salpeter equation. For simplicity, two functions
\begin{align}
\varphi_{_P}(\vec{q}) &\equiv i \int \frac{d q_{_0}}{2 \pi} \chi_{_P}(q),\\
\eta_{_P}(\vec{q}) &\equiv \int \frac{d\vec{k}}{(2 \pi)^{3}} V(\vec{k},\vec{q}) \varphi_{_P}(\vec{k}),
\end{align}
are defined. Thus, the BS equation can be rewritten as,
\begin{equation}\label{BS}
\chi_{_P}(q) = S_{1}(p_{1}) \eta_{_P}(\vec{q}) S_{2}(-p_{2}),
\end{equation}
where $S_{1}(p_{1})$ and $S_{2}(-p_{2})$ are the propagators of the quark 1 and antiquark 2, respectively.

For convenience, we write the formulas in covariant form. Therefore, we divide the relative momentum $q$ into two parts,  $q_{_{\parallel}}$ and  $q_{_{\perp}}$,
$$
q_{_{\parallel}} \equiv\frac {P\cdot q  }{M^{2}} P,~q_{_{\perp}} \equiv q-q_{_{\parallel}},$$
which are parallel and orthogonal to $P$ respectively, where $M$ is the mass of the relevant meson. Correspondingly, we have two Lorentz-invariant variables
$$
q_{_P}\equiv\frac{P \cdot q}{M},~q_{_{T}}\equiv\sqrt{-q_{_{\perp}}^{2}},
$$
which are $q_0$ and $|\vec q|$ respectively in the center of mass system.
The propagator $S_{i}$ ($i=1,2$) can be decomposed as,
\begin{equation}
S_{i} = \frac{\Lambda_{i P}^{+}(q_{_{\perp}})}{J(i) q_{_P}+\alpha_{i} M-\omega_{i }+i \epsilon} + \frac{\Lambda_{i P}^{-}(q_{_{\perp}})}{J(i) q_{_P}+\alpha_{i} M+\omega_{i }-i \epsilon},
\end{equation}
where $J(i) = (-1)^{i+1}$, $\omega_{i} = \sqrt{m_{i}^{2}+q_{_{T}}^{2}}$, and
$$
\Lambda_{i P}^{\pm}(q_{_{\perp}}) = \frac{1}{2 \omega_{i }}\left[\frac{\slashed P}{M} \omega_{i } \pm J(i)\left(m_{i}+\slashed q_{_{\perp}}\right)\right].
$$
The positive and negative energy projection operators $\Lambda^{+}$ and $\Lambda^{-}$ satisfy the following relations,
$$
\Lambda_{i P}^{+}(q_{_{\perp}}) + \Lambda_{i P}^{-}(q_{_{\perp}}) = \frac{\slashed P}{M}, ~
\Lambda_{i P}^{\pm}(q_{_{\perp}}) \frac{\slashed P}{M} \Lambda_{i P}^{\pm}(q_{_{\perp}}) = \Lambda_{i P}^{\pm}(q_{_{\perp}}), ~
\Lambda_{i P}^{\pm}(q_{_{\perp}}) \frac{\slashed P}{M} \Lambda_{i P}^{\mp}(q_{_{\perp}}) =0.
$$
Introducing the notation
\begin{equation}\label{ppmm}
\varphi_{_P}^{\pm\pm}(q_{_\perp}) \equiv \Lambda_{1 P}^{\pm}(q_{_\perp}) \frac{\slashed P}{M} \varphi_{_P}(q_{_\perp}) \frac{\slashed P}{M} \Lambda_{2 P}^{\pm}(q_{_\perp}),
\end{equation}
and using the relation $\frac{\slashed P}{M} \frac{\slashed P}{M} = 1$, we have:
\begin{equation}
\varphi_{_P}(q_{_\perp}) = \varphi_{_P}^{++}(q_{_\perp}) + \varphi_{_P}^{+-}(q_{_\perp}) + \varphi_{_P}^{-+}(q_{_\perp}) + \varphi_{_P}^{--}(q_{_\perp}).
\end{equation}

Further integrating out $q_{_0}$ on both sides of Eq.(\ref{BS}), we obtain the Salpeter equation,
\begin{equation}
\varphi_{_P}(q_{_\perp}) = \frac{\Lambda_{1 P}^{+}(q_{_\perp}) \eta_{_P}(q_{_\perp}) \Lambda_{2 P}^{+}(q_{_\perp})}{M - \omega_{1} - \omega_{2}} - \frac{\Lambda_{1 P}^{-}(q_{_\perp}) \eta_{_P}(q_{_\perp}) \Lambda_{2 P}^{-}(q_{_\perp})}{M + \omega_{1} + \omega_{2}}.
\end{equation}
By using the projection operators, this expression can be equivalently writen as,
\begin{align} \label{Lorentz-covariant}
(M - \omega_{1} - \omega_{2}) \varphi_{_P}^{++}(q_{_\perp}) &= \Lambda_{1 P}^{+}(q_{_\perp}) \eta_{_P}(q_{_\perp}) \Lambda_{2 P}^{+}(q_{_\perp}), \nonumber \\
(M + \omega_{1} + \omega_{2}) \varphi_{_P}^{--}(q_{_\perp}) &= -\Lambda_{1 P}^{-}(q_{_\perp}) \eta_{_P}(q_{_\perp}) \Lambda_{2 P}^{-}(q_{_\perp}), \\
\varphi_{_P}^{+-}(q_{_\perp}) &= \varphi_{_P}^{-+}(q_{_\perp}) = 0. \nonumber
\end{align}
The normalization condition for the Salpeter wave function is given by,
\begin{equation}\label{normalization condition}
\int \frac{d\vec{q}}{(2 \pi)^{3}} \operatorname{tr}\left[\bar{\varphi}_{_P}^{++}(q_{_\perp}) \frac{\slashed P}{M} \varphi_{_P}^{++}(q_{_\perp}) \frac{\slashed P}{M} - \bar{\varphi}_{_P}^{--} (q_{_\perp}) \frac{\slashed P}{M} \varphi_{_P}^{--}(q_{_\perp}) \frac{\slashed P}{M}\right] = 2 M,
\end{equation}
{where, $\bar{\varphi}=\gamma_0 \varphi^{\dag}\gamma^0$, '$\dag$' is the Hermitian conjugate transformation.} The relativistic BS equation is four-dimensional, but the Salpeter equation obtained through instantaneous approximation is three-dimensional. Therefore, strictly speaking, the Salpeter equation and the wave function obtained from it are semi-relativistic, not  completely relativistic.

In our previous works, the complete Salpeter equations for the pseudoscalar \cite{wang0-} and tensor \cite{wang2+} mesons have been solved. The Cornell potential is chosen as the interaction kernel: $$V(r)=\lambda r+V_{0}-\gamma_{0}\otimes\gamma^{0}\frac{4}{3}\frac{\alpha_{s}(r)}{r},$$ where $\lambda$ is the string constant (for heavy-light mesons, $\lambda=0.25$ GeV$^{2}$), $\alpha_{s}(r)$ is the running coupling constant, and $V_{0}$ is a free constant. The interaction potential in momentum space is given by,
\begin{eqnarray}
V(\vec{q}) &=& V_s(\vec{q})+\gamma_0\otimes\gamma^0V_\upsilon(\vec{q}), \\
V_s(\vec{q}) &=& -\Big(\frac{\lambda}{\alpha}+V_0\Big)\delta^3(\vec{q})+\frac{\lambda}{\pi^2}\frac{1}{(\vec{q}^2+\alpha^2)^2}, \\
V_\upsilon(\vec{q}) &=& -\frac{2}{3\pi^2}\frac{\alpha_s(\vec{q})}{(\vec{q}^2+\alpha^2)}, \end{eqnarray}
where $$\alpha_s(\vec{q}) = \frac{12\pi}{27}\frac{1}{\log(a+\frac{\vec{q}^2}{\Lambda_{QCD}^2})},$$ $a=2.7183$ and $\Lambda_{QCD}=0.27$ GeV.
In order to avoid infrared divergence and incorporate the screening effect, a small  parameter $\alpha=0.06$ GeV is added in the potential.}

\section{ Wave functions and their partial waves.}
\subsection{$0^{-}$ meson}
{
Usually, people do not solve the complete Salpeter equation, namely the four equations of Eq.(\ref{Lorentz-covariant}), but only solve the first one, which is about the positive energy wave function. Due to the fact that $(M + \omega_{1} + \omega_{2})$$\gg$$(M - \omega_{1} - \omega_{2})$, we have $\varphi_{_P}^{++}(q_{_\perp})$$\gg$$\varphi_{_P}^{--}(q_{_\perp})$, and $\varphi_{_P}^{--}(q_{_\perp})$ is negligible. However, this approach also ignored most of the relativistic corrections, because one equation can only solve the case with only one unknown radial wave function. For example, the wave function
\begin{equation}\label{non}\varphi_{_P}(q_{_\perp})=(\frac{\slashed P}{M}+1)\gamma^{5}f(q_{_\perp})
\end{equation}  is for a pseudoscalar,
where the radial wave function $f(q_{_\perp})\equiv f(-q^2_{_\perp})$ can be obtained numerically by solving the first equation in Eq.(\ref{Lorentz-covariant}). Due to the absence of standalone $q_{_\perp}$ terms, this solving `incomplete Salpeter equation' method can only obtain non-relativistic wave function rather than a relativistic one.
The correct and safe way is to first solve the complete Salpeter equation and obtain the positive and negative energy wave functions, and then omit the contribution of the negative energy wave function in specific applications.

The Salpeter wave function for a $0^{-}$ state has the general form \cite{wang0-}
\begin{equation}\label{0- wave}
\varphi _{0^-}\left(q_{\bot }\right)=\left[\frac{\slashed{P}}{M}f_1\left(q_{\bot }\right)+f_2\left(q_{\bot }\right)+\frac{ \slashed{q}_{\bot }}{M}f_3\left(q_{\bot }\right)+\frac{\slashed{P} \slashed{q}_{\bot }}{M^2}f_4\left(q_{\bot }\right)\right]\gamma ^5.
\end{equation}
We have four unknown radial wave functions $f_i's$, which are function of $-q^2_{_\perp}$. Compared with the non-relativistic wave function in Eq.(\ref{non}), our Salpeter wave function has two additional relativistic terms, namely $f_3$ and $f_4$ terms, and $f_1\neq f_2$. So it contains rich relativistic information. Using the last two equations of Salpeter Eq.(\ref{Lorentz-covariant}), we get
\begin{eqnarray}
&&f_3\left(q_{\bot }\right)=\frac{f_2 M \left(\omega _2-\omega _1\right)}{m_2 \omega _1+m_1 \omega _2},
\quad f_4\left(q_{\bot }\right)=-\frac{f_1 M \left(\omega _1+\omega _2\right)}{m_2 \omega _1+m_1 \omega _2} \nonumber.
\end{eqnarray}
By using the first two equations of the Salpeter Eq.(\ref{Lorentz-covariant}), the two unknown independent radial wave functions $f_1$ an $f_2$ will be obtained. We do not give the detailed calculation for the $0^-$ state here. Instead, we will provide the detailed calculation for the more complex $2^+$ state in the next subsection.
The normalization condition Eq.(\ref{normalization condition}) for this $0^-$ wave function is \cite{wang0-}
\begin{equation}\label{0-normal}
\int \frac{d\vec q}{(2\pi)^3} \frac{8 M \omega_1 \omega_2 f_1 f_2}{\omega_1 m_2 + \omega_2 m_1}=1.
\end{equation}
}

We have pointed out that the wave function of the $0^{-}$ state not only contains $S$-wave, namely the terms with $f_1$ and $f_2$, but also $P$-wave components, namely $f_3$ and $f_4$ terms \cite{part wave}. {If we only consider the contribution of  $S-$wave, the normalization formula Eq.(\ref{0-normal}) is}
\begin{equation}\label{0-Snormal}
\int \frac{d\vec q}{(2\pi)^3} \frac{2 M f_1 f_2 (\omega_1 m_2 + \omega_2 m_1)}{\omega_1 \omega_2}.
\end{equation}
Based on Eq.(\ref{0-normal}) and Eq.(\ref{0-Snormal}), which are $(S+P)^2$ and $S^2$, the ratio of $S$ partial wave and $P$ wave can be calculated \cite{part wave}.

For the $0^{-}$ meson, its positive energy wave function, Eq.(\ref{ppmm}), can be expressed as
\begin{equation}
\varphi _{0^-}^{++}\left(q_{\bot }\right)=\left[A_1\left(q_{\bot }\right)+\frac{\slashed{P}}{M}A_2\left(q_{\bot }\right)+\frac{\slashed{q}_{\bot }}{M}A_3\left(q_{\bot }\right)+\frac{\slashed{P} \slashed{q}_{\bot }}{M^2}A_4\left(q_{\bot }\right)\right]\gamma ^5,
\end{equation}
where $A_1$ and $A_2$ terms are $S$ waves, and $A_3$ and $A_4$ terms are $P$ wave. Their detailed expressions are as follows,
\begin{eqnarray}
   && A_1=\frac{M}{2}\left(\frac{f_1 \left(\omega _1+\omega _2\right)}{m_1+m_2}+f_2\right),~~A_3=-\frac{A_1 M \left(\omega _1-\omega _2\right)}{m_2 \omega_1+m_1 \omega_2},  \nonumber
 \\ &&A_2=\frac{M}{2}\left(\frac{f_2 \left(m_1+m_2\right)}{\omega _1+\omega _2}+f_1\right),~~  A_4=-\frac{A_1 M \left(m_1+m_2\right)}{m_2 \omega_1+m_1 \omega_2}. \nonumber
\end{eqnarray}
\subsection{$2^{+}$ Meson}
{The bound state with quantum number $J^{P} = 2^{+}$ can be described by the following Salpeter wave function \cite{wang2+},
\begin{eqnarray}\label{2++ wave function}
\varphi _{2^{+}}\left(q_{_\perp }\right)=\epsilon _{\mu\nu}q_{_\perp }^{\mu }q_{_\perp }^{\nu}\left[\zeta_1\left(q_{_\perp }\right)+\frac{\slashed{P}}{M}\zeta_2\left(q_{_\perp }\right)+\frac{\slashed{q}_{_\perp }}{M}\zeta_3\left(q_{_\perp }\right)+\frac{\slashed{P}\slashed{q}_{_\perp }}{M^2}\zeta_4\left(q_{_\perp }\right)\right] \nonumber
\\+M\epsilon _{\mu\nu}\gamma ^{\mu }q_{_\perp }^{\nu}\left[\zeta_5\left(q_{_\perp }\right)+\frac{\slashed{P}}{M}\zeta_6\left(q_{_\perp }\right)+\frac{\slashed{q}_{_\perp }}{M}\zeta_7\left(q_{_\perp }\right)+\frac{\slashed{P}\slashed{q}_{_\perp }}{M^2}\zeta_8\left(q_{_\perp }\right)\right],
\end{eqnarray}
where $\varepsilon_{\mu\nu}$ is the symmetric polarization tensor of the meson; the unknown radial wave function $\zeta_i(q_{_\perp })$ ($i=1,2...8$) is function of $-q^2_{_\perp }$, which will be obtained by solving the Salpeter equation. Using the last two equations of Eq.(\ref{Lorentz-covariant}), it is found that only four radial wave functions are independent. We choose $\zeta_3(q_{_\perp })$, $\zeta_4(q_{_\perp })$, $\zeta_5(q_{_\perp })$, and $\zeta_6(q_{_\perp })$ as the independent ones, and others can be expressed as
\begin{eqnarray}
 && \zeta _1\left(q_{_\perp }\right)=\frac{q_{_\perp }^2\zeta _3 \left(\omega _1+\omega _2\right)+2 \zeta _5 M^2 \omega _2}{M \left(m_2 \omega _1+m_1 \omega _2\right)},
 ~~\zeta _7\left(q_{_\perp }\right)=\frac{M \left(\omega _1-\omega _2\right)}{m_2 \omega _1+m_1 \omega _2} \zeta _5 ,\nonumber \\
&& \zeta _2\left(q_{_\perp }\right)=\frac{q_{_\perp }^2\zeta _4 \left(\omega _1-\omega _2\right)+2 \zeta _6 M^2 \omega _2}{M \left(m_2 \omega _1+m_1 \omega _2\right)}, ~~
\zeta _8\left(q_{_\perp }\right)=\frac{M \left(\omega _1+\omega _2\right)}{m_2 \omega _1+m_1 \omega _2} \zeta _6.\nonumber
\end{eqnarray}

From Eq.(\ref{ppmm}), we obtain the expressions of positive and negative energy wave functions. For example, the positive energy wave function of $2^{+}$ meson is
\begin{eqnarray}\label{2+ postive wave function}
\varphi _{2^{+}}^{++}\left(q_{\bot }\right)=\epsilon _{\mu\nu}q_{\bot }^{\mu }q_{\bot }^{\nu}\left[B_1\left(q_{\bot }\right)+\frac{\slashed{P}}{M}B_2\left(q_{\bot }\right)+\frac{\slashed{q}_{\bot }}{M}B_3\left(q_{\bot }\right)+\frac{\slashed{P}\slashed{q}_{\bot }}{M^2}B_4\left(q_{\bot }\right)\right] \nonumber
\\+M\epsilon _{\mu\nu}\gamma ^{\mu }q_{\bot }^{\nu}\left[B_5\left(q_{\bot }\right)+\frac{\slashed{P}}{M}B_6\left(q_{\bot }\right)+\frac{\slashed{q}_{\bot }}{M}B_7\left(q_{\bot }\right)+\frac{\slashed{P}\slashed{q}_{\bot }}{M^2}B_8\left(q_{\bot }\right)\right],
\end{eqnarray}
where $B_i$s are functions of four independent radial wave functions $\zeta _3,\zeta _4,\zeta _5,$ and $\zeta _6$. Their specific expression are denoted as
\begin{eqnarray}
B_{1} && =  \frac{1}{2 M\left(m_{1} \omega_{2}+m_{2} \omega_{1}\right)}\left[\left(\omega_{1}+\omega_{2}\right) q_{_\perp}^{2} \zeta_{3}+\left(m_{1}+m_{2}\right) q_{_\perp}^{2} \zeta_{4}+2 M^{2} \omega_{2} \zeta_{5}-2 M^{2} m_{2} \zeta_{6}\right], \nonumber \\
B_{2} && =  \frac{1}{2 M\left(m_{1} \omega_{2}+m_{2} \omega_{1}\right)}\left[\left(m_{1}-m_{2}\right) q_{_\perp}^{2} \zeta_{3}+\left(\omega_{1}-\omega_{2}\right) q_{_\perp}^{2} \zeta_{4}+2 M^{2} \omega_{2} \zeta_{6}-2 M^{2} m_{2} \zeta_{5}\right] , \nonumber\\
B_{3} && =  \frac{1}{2}\left[\zeta_{3}+\frac{m_{1}+m_{2}}{\omega_{1}+\omega_{2}} \zeta_{4}-\frac{2 M^{2}}{m_{1} \omega_{2}+m_{2} \omega_{1}} \zeta_{6}\right],~~B_{5} =  \frac{1}{2}\left[\zeta_{5}-\frac{\omega_{1}+\omega_{2}}{m_{1}+m_{2}} \zeta_{6}\right], \nonumber \\
B_{4} && =  \frac{1}{2}\left[\frac{\omega_{1}+\omega_{2}}{m_{1}+m_{2}} \zeta_{3}+\zeta_{4}-\frac{2 M^{2}}{m_{1} \omega_{2}+m_{2} \omega_{1}} \zeta_{5}\right] ,~~ B_{6} =  \frac{1}{2}\left[-\frac{m_{1}+m_{2}}{\omega_{1}+\omega_{2}} \zeta_{5}+\zeta_{6}\right] , \nonumber\\
B_{7} && =  \frac{M}{2} \frac{\omega_{1}-\omega_{2}}{m_{1} \omega_{2}+m_{2} \omega_{1}}\left[\zeta_{5}-\frac{\omega_{1}+\omega_{2}}{m_{1}+m_{2}} \zeta_{6}\right], ~
B_{8} =  \frac{M}{2} \frac{m_{1}+m_{2}}{m_{1} \omega_{2}+m_{2} \omega_{1}}\left[-\zeta_{5}+\frac{\omega_{1}+\omega_{2}}{m_{1}+m_{2}} \zeta_{6}\right]. \nonumber
\end{eqnarray}

Substituting the positive and negative wave functions into the first two equations of Eq.(\ref{Lorentz-covariant}), and multiplying both sides by the same variable (for example, $\slashed{P}$, $\slashed{q}_{_\perp}$, $\slashed{q}_{_\perp}\slashed{P}$, etc), we calculate the trace on both sides, and find there are four (not two) independent eigenvalue equations. For convenience, we have replaced $\zeta_3(q_{_\perp})$, $\zeta_4(q_{_\perp})$, $\zeta_5(q_{_\perp})$, $\zeta_6(q_{_\perp})$ with $F_1(q_{_\perp})$, $F_2(q_{_\perp})$, $F_3(q_{_\perp})$, $F_4(q_{_\perp})$. Their relations are
\begin{align}\label{F1234}
F_1 (q_{_\perp})&= \frac{4q_{_\perp}^4[(\omega_1 + \omega_2)(\zeta_5M^2 + \zeta_3q_{_\perp}^2) - (m_1 + m_2)(\zeta_6M^2 - \zeta_4q_{_\perp}^2)]}{3M(m_2\omega_1 + m_1\omega_2)}, \\
F_2 (q_{_\perp})&= \frac{4q_{_\perp}^4[(m_1 + m_2)(\zeta_6M^2 - \zeta_4q_{_\perp}^2) + (\omega_1 + \omega_2)(\zeta_5M^2 + \zeta_3q_{_\perp}^2)]}{3M(m_2\omega_1 + m_1\omega_2)}, \\
F_3 (q_{_\perp})&= \frac{2q_{_\perp}^4[-\zeta_5(5m_1 + m_2)M^2 - 2(\zeta_3(m_1 - m_2)q_{_\perp}^2 + \zeta_4q_{_\perp}^2(\omega_1 - \omega_2)) + \zeta_6M^2(5\omega_1 + \omega_2)]}{3M(m_2\omega_1 + m_1\omega_2)}, \\
F_4 (q_{_\perp})&= \frac{2q_{_\perp}^4[\zeta_5(5m_1 + m_2)M^2 - 2(\zeta_4q_{_\perp}^2(\omega_1 - \omega_2) - \zeta_3(m_1 - m_2)q_{_\perp}^2) + \zeta_6M^2(5\omega_1 + \omega_2)]}{3M(m_2\omega_1 + m_1\omega_2)}.
\end{align}
The obtained four coupled eigenvalue equations are presented in the Appendix.

The normalization condition Eq.(\ref{normalization condition}) for the $2^+$  wave function is \cite{wang2+}
\begin{equation}\label{2++ norm}
\int \frac{d\vec q}{(2\pi)^3} \frac{8M \omega_1 \omega_2 \vec{q}^2}{15(\omega _1m_2+\omega_2 m_1)}
[-\zeta_5\zeta_6+\frac{2\vec{q}^2}{M^2}(-\zeta_4 \zeta_5+\zeta_3 \zeta_6+\zeta_3\zeta_4\frac{\vec{q}^2 }{M^2} ) ]=1.
\end{equation}}

In our expression, the $2^{+}$ state $\bar D_{2}^\star(2460)^0$, is not a pure $P$-wave, but contains both $D$ and $F$ partial waves \cite{part wave}. In Eq.(\ref{2++ wave function}), {the terms including $\zeta_5$ and $\zeta_6$ are $P$ waves which are non-relativistic,} $\zeta_3$ and $\zeta_4$ terms are $F-P$ mixing waves, and others are $D$ waves. Thus, we can conclude that the wave function of the tensor $\bar D^{\star}_{2}(2460)^0$ contains $P$, $D$, and $F$ partial waves.

If only the pure $P$ wave is considered, the wave function of the $2^{+}$ meson becomes
\begin{eqnarray}\label{2++P}
\varphi _{2^{+}}^{P}\left(q_{\bot }\right)=\epsilon _{\mu\nu}q_{\bot }^{\mu }\gamma^{\nu}(M \zeta_5+\slashed P \zeta_6)+\frac{2}{5}
\epsilon _{\mu\nu}q_{\bot }^{\mu }\gamma^{\nu}q_{\bot }^{2 }(\frac{\zeta_3}{M}-\frac{\slashed P}{M^2} \zeta_4 ),
\end{eqnarray}
{and the contribution of this $P$ partial wave to the overall normalization condition Eq.(\ref{2++ norm}) is}
\begin{eqnarray}\label{2++P norm}
\int \frac{d\vec q}{(2\pi)^3} \frac{2 \vec{q}^2 \left(2 \zeta_{3} \vec{q}^2-5 \zeta_{5} M^2\right) \left(2 \zeta_{4} \vec{q}^2+5 \zeta_{6} M^2\right)(\omega _1 m_2 +\omega _2 m_1 )}{75 M^3 \omega _1 \omega _2}.
\end{eqnarray}
While for a pure $F$ wave, the wave function is
\begin{eqnarray}
\varphi _{2^{+}}^{F}\left(q_{\bot }\right)=\epsilon _{\mu\nu}q_{\bot }^{\mu }q_{\bot }^{\nu}
(\frac{\slashed {q}_\bot}{M}\zeta_3+\frac{\slashed P \slashed {q}_\bot  }{M^2} \zeta_4)-\frac{2}{5}
\epsilon _{\mu\nu}q_{\bot }^{\mu }\gamma^{\nu} q_{\bot }^{2 }(\frac{\zeta_3}{M}-\frac{\slashed P}{M^2} \zeta_4 ),
\end{eqnarray}
{and its contribution to the normalization condition Eq.(\ref{2++ norm}) is}
\begin{equation}\label{2++F norm}
 \int \frac{d\vec q}{(2\pi)^3} \frac{4 \zeta_3 \zeta_4 \vec {q} ^6(\omega _1 m_2 +\omega _2 m_1 )}{25 M^3 \omega _1 \omega _2}.
\end{equation}
Using Eq.(\ref{2++ norm}), Eq.(\ref{2++P norm}), and Eq.(\ref{2++F norm}), we can calculate the ratios between different partial waves.

\section{semileptonic decay width formula}
For the $B^+ \to \bar D_{2}^\star(2460)^0 \ell^+ \nu_\ell$ process as shown in Fig. 1, the transition amplitude is written as
\begin{eqnarray}
T = \frac{G_F}{\sqrt{2}}V_{cb}\bar{\mu}_{\nu_\ell}  \gamma ^{\mu}(1-\gamma ^5)\upsilon _{\ell}
\left \langle  D_{2}^\star(2460)^0(P_f) |J_\mu |B^+(P)  \right \rangle ,
\end{eqnarray}
where $G_{F}$ is the Fermi constant, $J_{\mu} \equiv V_{\mu} - A_{\mu}$ is the charged current responsible for the decays, $V_{cb} = 40.5 \times 10^{-3}$ (PDG \cite{PDG}) is CKM matrix element, { $\mu_{\nu_\ell}$ is the spinor of the neutrino $\nu_\ell$, $\upsilon _{\ell}$ is the spinor of the anti-lepton $\ell^+$,} $P$ and $P_f$ are the momenta of the initial $B^+$ and the final $\bar D_{2}^\star(2460)^0$, respectively.
\begin{figure}[H]
\begin{picture}(270,200)(110,540)
\put(0,0){\includegraphics{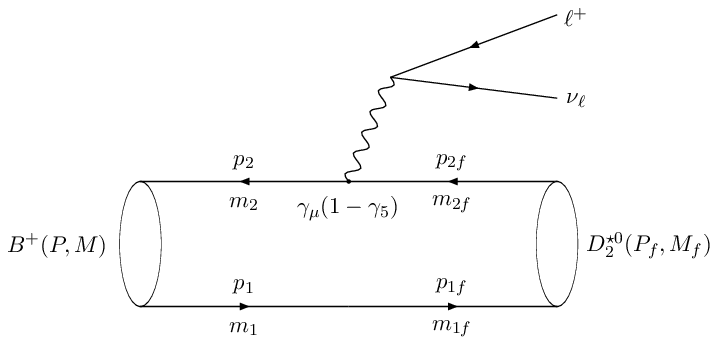}}
\end{picture}
\caption{Feynman diagram corresponding to the semileptonic decay $B^+ \to \bar D_{2}^\star(2460)^0 \ell^+ \nu_\ell$
}\label{figure1}
\end{figure}
After summing the polarizations of the initial and final mesons, the square of the above matrix elements is written as
\begin{eqnarray}
\sum \left | T^2 \right | = \frac{G_F^2}{2} \left |  V_{cb}\right | ^2 \ell^{\mu \nu}h_{\mu \nu},
\end{eqnarray}
where $\ell^{\mu \nu} \equiv {\sum}\bar{\mu}_{\nu_\ell}\gamma^\mu(1-\gamma_5)\upsilon _\ell\bar{\upsilon }_\ell(1+\gamma_5)\gamma^\nu\mu_{\nu_\ell}$ is the leptonic tensor, and $h_{\mu \nu} \equiv {\sum}\left \langle B^+(P) |J_\nu^+ | D_{2}^\star(2460)^0 (P_f) \right \rangle \left \langle D_{2}^\star(2460)^0 (P_f) |J_\mu |B^+(P)  \right \rangle$ is the hadron tensor.

By using Mandelstam's formulism, the hadronic transition matrix element can be written as the overlapping integral over the BS wave functions of initial and final mesons. Since we do not solve the BS equation, but the Salpeter equation, the transition matrix element is further simplified by instantaneous approximation. Then for the process $B^+ \to \bar D_{2}^\star(2460)^0 \ell^+ \nu_\ell$, the hadronic matrix element can be written as \cite{changwang}
\begin{eqnarray}
 && \left \langle  D_{2}^\star(2460)^0 (P_f) |J_\mu |B^+(P)  \right \rangle =\int\frac{d\vec{q}}{(2 \pi )^3} \mathrm{Tr}\left[\frac{\slashed P}{M} \varphi^{++}_P(\vec{q}) \gamma _\mu (1-\gamma _5)  \bar{\varphi}^{++}_{P_f} (\vec{q_f}) \right]\label{transi}  \nonumber \\&&=t_1~\epsilon_{\mu P}+t_2~\epsilon_{P P} P_\mu +t_3~\epsilon_{P P} P_{f \mu}  +i~t_4~ \epsilon^{\rho P} \varepsilon_{\rho P P_f \mu},
\end{eqnarray}
where {$\vec q_{f} =\vec q - \alpha_{1f}\vec P_{f}$ is used, which is obtained by assuming the momentum of spectator quark remains unchanged;} $\epsilon_{\mu \nu}$ is the polarization tensor of the final tensor meson; $t_1$, $t_2$, $t_3$ and $t_4$ are the form factors; $\varphi^{++}_P$ and {$\bar{\varphi}^{++}_{P_f}=\gamma_0({{\varphi}^{++}_{P_f} })^\dagger\gamma^0$} are the positive energy Salpeter wave functions for the initial and final mesons, respectively. We have used the abbreviations, for example $\epsilon^{\rho\sigma} {P_{\sigma}}~ \varepsilon_{\rho\alpha\beta  \mu}P^{\alpha} {P_f}^{\beta}=\epsilon^{\rho P} \varepsilon_{\rho P P_f \mu}$.

Based on the covariance analysis of the Lorenz index, the general form of $h_{\mu \nu}$ can be expressed as
\begin{eqnarray}
h_{\mu \nu} = &&-\alpha g_{\mu \nu}+\beta_{++}\left(P+P_f\right)_{\mu}\left(P+P_f\right)_{\nu}+\beta_{+-}\left(P+P_f\right)_{\mu}\left(P-P_f\right)_{\nu} \nonumber \\
&&+\beta_{-+}\left(P-P_f\right)_{\mu}\left(P+P_f\right)_{\nu}+\beta_{--}\left(P-P_f\right)_{\mu}\left(P-P_f\right)_{\nu} \nonumber \\
&&+ i \gamma \varepsilon_{\mu \nu \rho \sigma}\left(P+P_f\right)^{\rho}\left(P-P_f\right)^{\sigma},
\end{eqnarray}
where, the coefficients $\alpha$, $\beta_{\pm\pm}$ and $\gamma$ are functions of the form factors $t_i~ (i=1,2,3,4)$.
Thus, the differential decay rate of this { exclusive process} can be written as
\begin{eqnarray}
\frac{d^{2} \Gamma}{d x d y}= && \left|V_{i j}\right|^{2} \frac{G_{F}^{2} M^{5}}{32 \pi^{3}}\left\{\alpha \frac{\left(y-\frac{m_{\ell}^{2}}{M^{2}}\right)}{M^{2}}+2 \beta_{++} \right. \nonumber \\
&& \times{\left[2 x\left(1-\frac{M_f^{ 2}}{M^{2}}+y\right)-4 x^{2}-y+\frac{m_{\ell}^{2}}{4 M^{2}}\left(8 x+\frac{4 M_f^{2}-m_{\ell}^{2}}{M^{2}}-3 y\right)\right]} \nonumber \\
&& +\left(\beta_{+-}+\beta_{-+}\right) \frac{m_{\ell}^{2}}{M^{2}}\left(2-4 x+y-\frac{2 M_f^{2}-m_{\ell}^{2}}{M^{2}}\right)+ \beta_{--} \frac{m_{\ell}^{2}}{M^{2}}\left(y-\frac{m_{\ell}^{2}}{M^{2}}\right) \nonumber \\
&& \left.-\gamma\left[y\left(1-\frac{M_f^{2}}{M^{2}}-4 x+y\right)+\frac{m_{\ell}^{2}}{M^{2}}\left(1-\frac{M_f^{2}}{M^{2}}+y\right)\right]\right\},
\end{eqnarray}
where $x\equiv E_\ell/M,~y\equiv (P-P_f)^2/M^2$, $M$ and $M_f$ are the masses of $B^+$ and $\bar D_{2}^\star (2460)^0$, respectively, and {$m_\ell$ and $E_\ell$} are the mass and energy of the final charged lepton $\ell$, respectively.

\section{Results and discussion}
In the calculation, we used the constituent quark masses: $m_u=0.38$ GeV, $m_d=0.385$ GeV, $m_s=0.55$ GeV, $m_c=1.62$ GeV, and $m_b=4.96$ GeV. Other model-dependent parameters have been shown in the text.
{In our calculation, for each $J^P$, the ground state mass is our input,  that is, its value is obtained by fitting experimental data to determine the free parameter $V_0$. And the masses of excited states are our predictions. Therefore, the masses of $B$, $B_s$, and $B_c$ are the same as the experimental values and are not reiterated here. The mass spectra for the tensor $2^{++}$ states are provided in Table \ref{mass}.}

\begin{table*}[hbt]
\caption{Mass spectra of the $2^+$ tensors in unit of GeV.}\label{mass}
\begin{tabular*}{\textwidth}{@{}c@{\extracolsep{\fill}}ccccc}
\hline \hline
&$m_{\bar D_2^{\star 0}}$&$m_{D_2^{\star -}}$&$m_{D_{s2}^{\star -}}$&$m_{\chi_{c2}}$\\ %\hline
$1~^3P_2$ &2.461~{(input)}&2.465~{(input)}&2.569~{(input)}&3.556~{(input)}
\\
$2~^3P_2$ &2.985&2.992&3.111&3.972
\\
$3~^3P_2$ &3.342&3.352&3.474&4.270
\\
\hline\hline
\end{tabular*}
\end{table*}

\subsection{The wave functions and ratios of different partial waves}

\begin{figure}[h!]
    \centering
    \includegraphics[width=6cm]{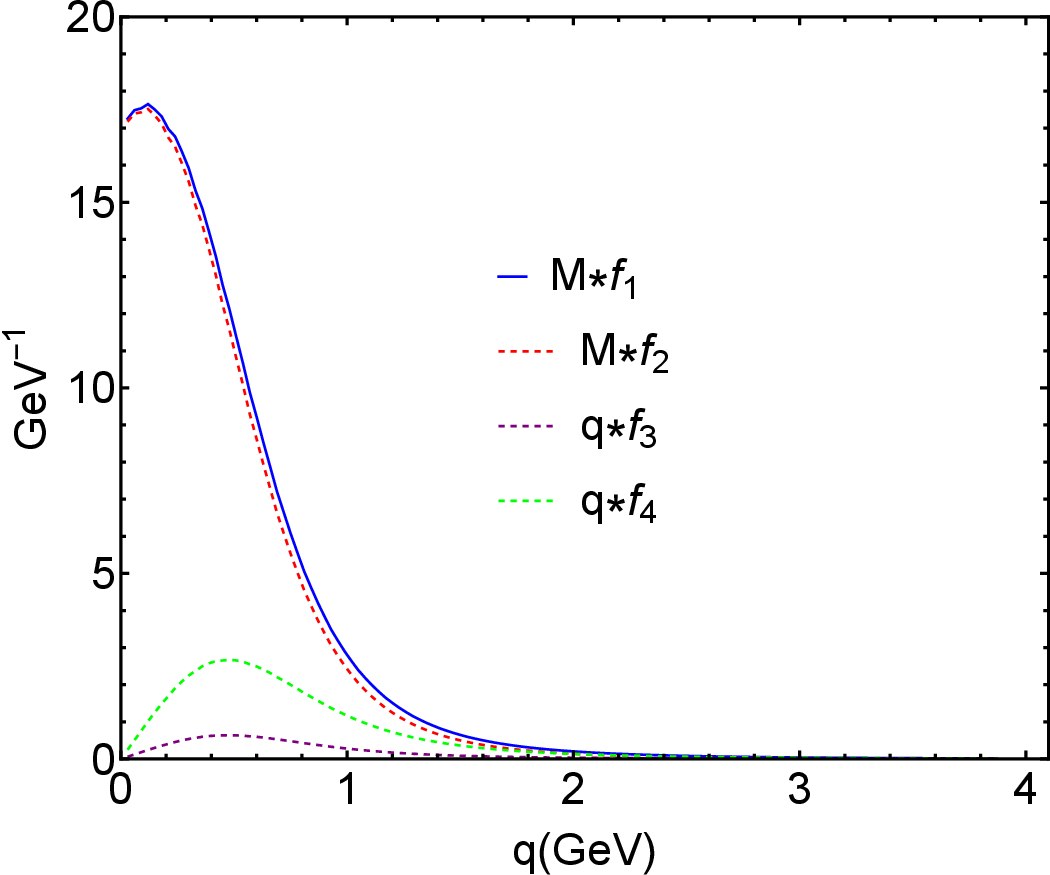}
    \includegraphics[width=6cm]{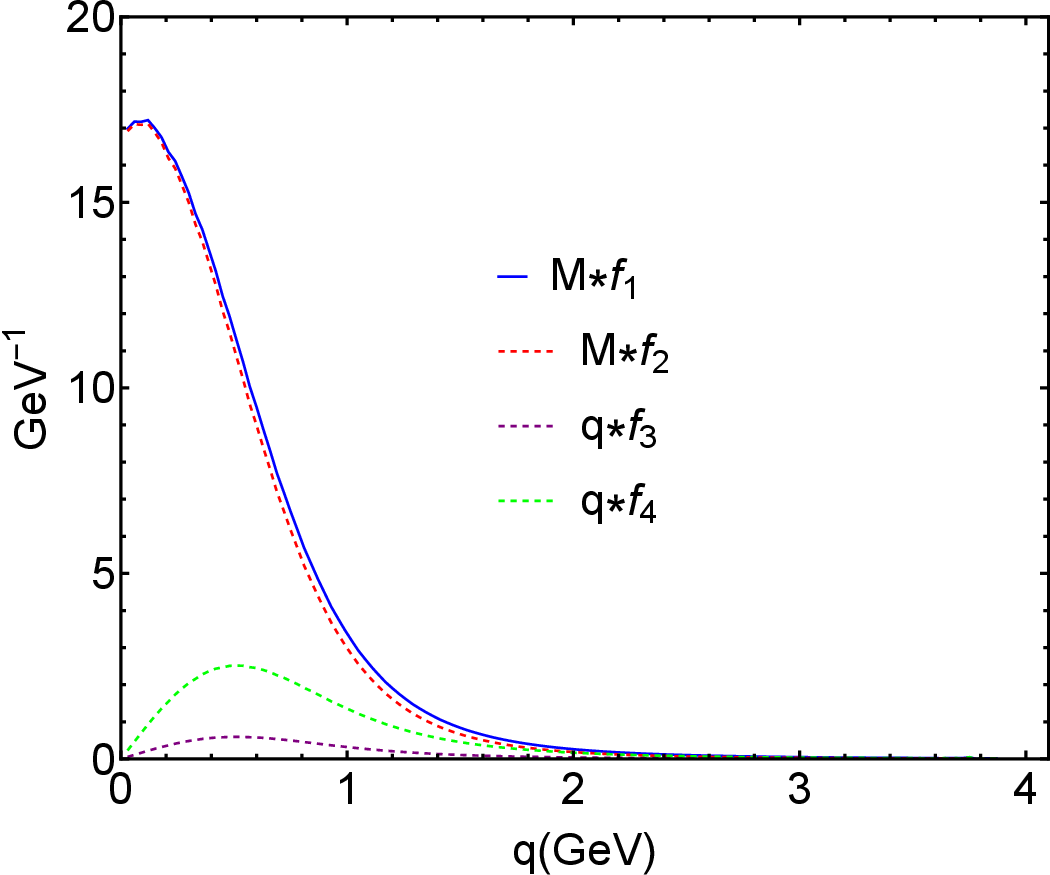}
    \centering \caption{The radial wave functions of the $0^{-}$ mesons $B^+$ (left) and $B_s^0$ (right), where $f_1$ and $f_2$ terms are $S$ waves;  $f_3$ and $f_4$ terms are $P$ waves, and $q\equiv|\vec q|$.} \label{B+ and Bs}
\end{figure}

The numerical values of the radial wave functions for $0^-$ pseudoscalars $B$ and $B_s$ are shown in Figure \ref{B+ and Bs} (For $B_c$ meson, see Ref.\cite{part wave}). We can see the $S$-wave components, namely the $f_1$ and $f_2$ terms in Eq.(\ref{0- wave}), are dominant, and the $P$-wave ones, namely $f_3$ and $f_4$ terms, are small. So, $B^+$, $B^0$, $B_s^0$ and $B_c^+$ are all $S$ wave dominant states. To see this clearly, we calculate the ratio between $S$ and $P$ waves which are based on the normalization formulas Eq.(\ref{0-normal}) and Eq.(\ref{0-Snormal}), and the results are shown in Table \ref{0- partial wave}.

In a non-relativistic limit, only the $S$ wave survives, and $f_1=f_2$ for a $0^-$ meson. In our {semi-relativistic method}, first, radial wave function $f_1$ is not exactly equal to $f_2$; second, the $0^-$ wave function also includes the $P$ wave components, $f_3$ and $f_4$ terms, which contribute to the relativistic correction. The ratios in Table \ref{0- partial wave}, indicate that the relativistic correction in $B^+$ (or $B^0$) is a little larger than that of $B_s^0$, and much larger than that of $B_c^+$.
\begin{table}[ht]
\centering \caption{Ratios between the $S$ wave and $P$ wave in the $0^{-}$ wave function.} \label{0- partial wave}
\setlength{\tabcolsep}{6pt}
\renewcommand{\arraystretch}{1}
\begin{tabular}{|c|c|c|c|c|c|}
\hline
   $0^{-}$ meson &     $B^+$   &    $B^0$    &  $B_s^0$   &  $B_c^+$   \\ \hline
    $S:P$    &   1~:~0.339  &  1~:~0.333    &  1~:~0.227   &  1~:~0.0815  \\ \hline
\end{tabular}
\end{table}

\begin{figure}[h!]
    \centering
    \includegraphics[width=5.3cm]{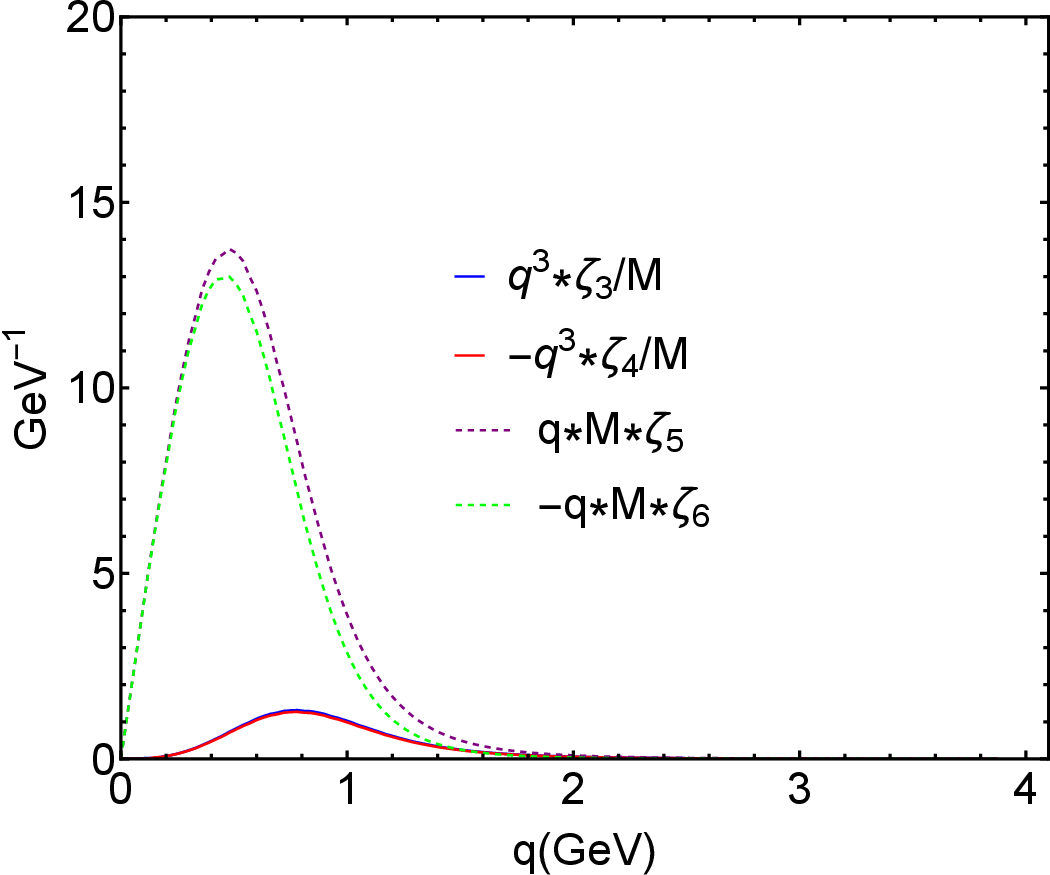}
    \includegraphics[width=5.3cm]{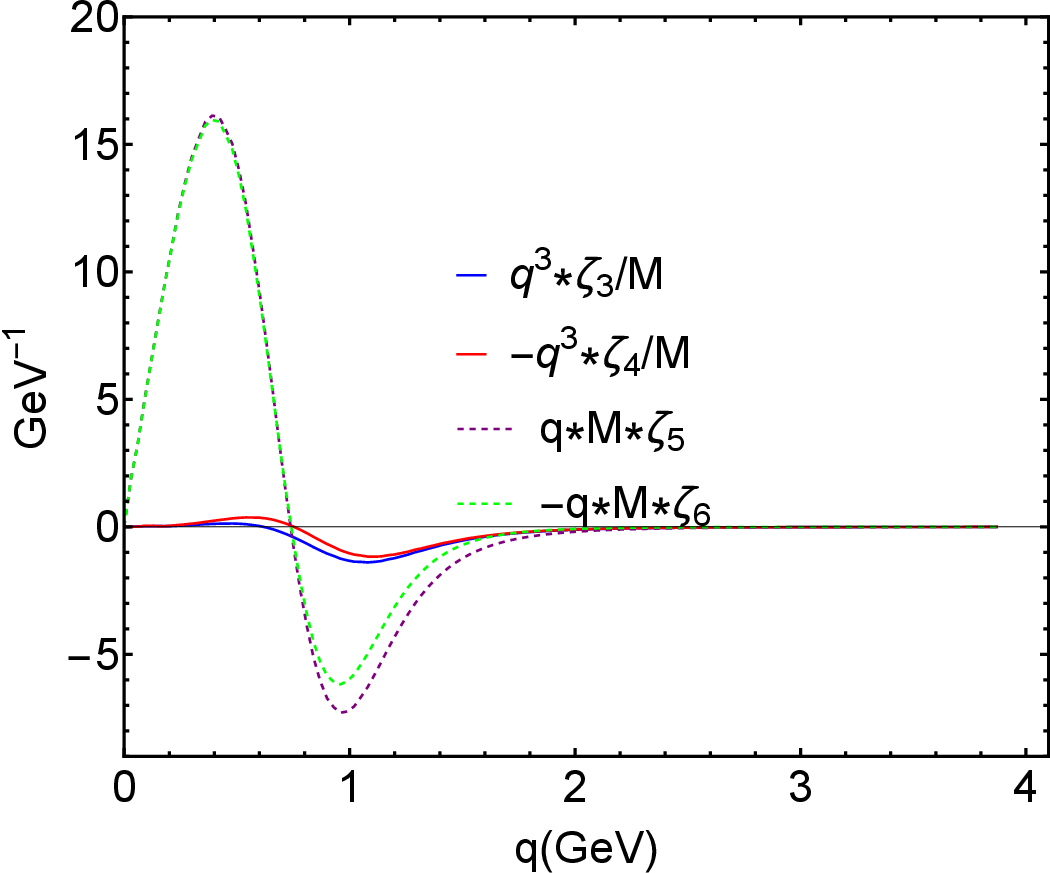}
    \includegraphics[width=5.3cm]{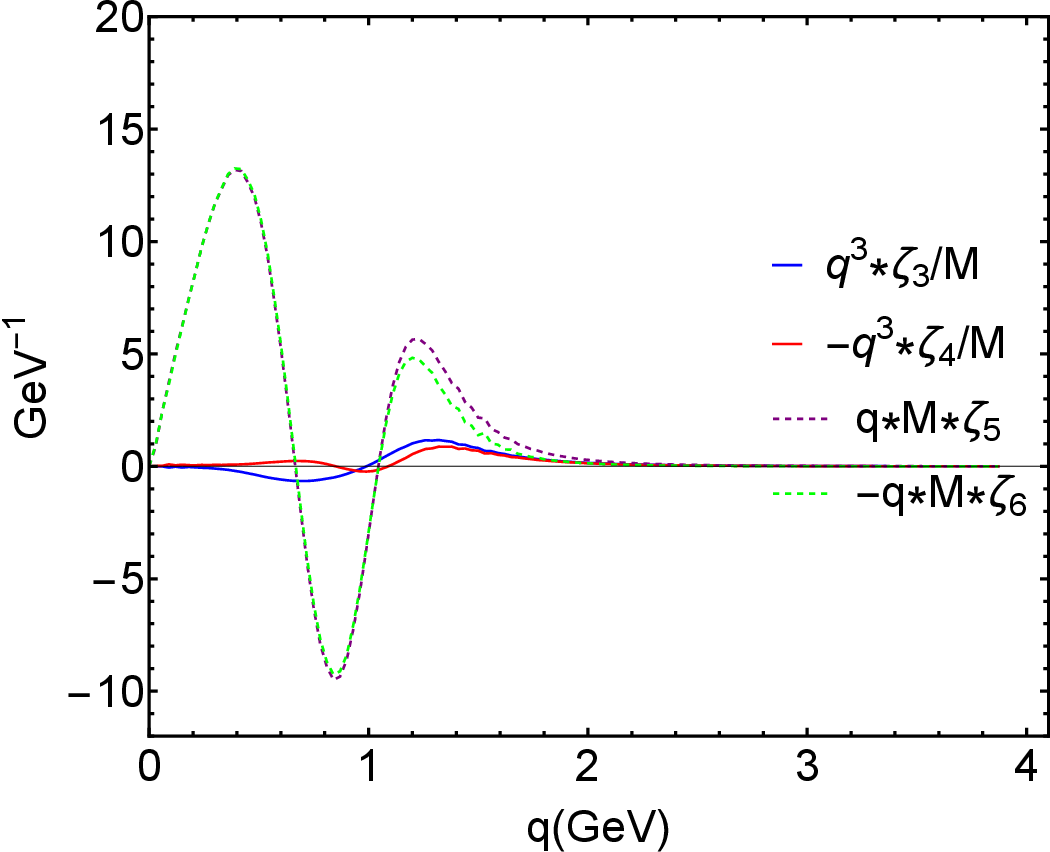}
    \centering \caption{The four independent radial wave functions for the $2^+$ tensors $\bar D_2^{\star 0}(1P)$  (left), $\bar D_2^{\star 0}(2P)$ (middle) and the $\bar D_2^{\star 0}(3P)$ (right).} \label{D20}
\end{figure}

\begin{figure}[h!]
    \centering
    \includegraphics[width=5.3cm]{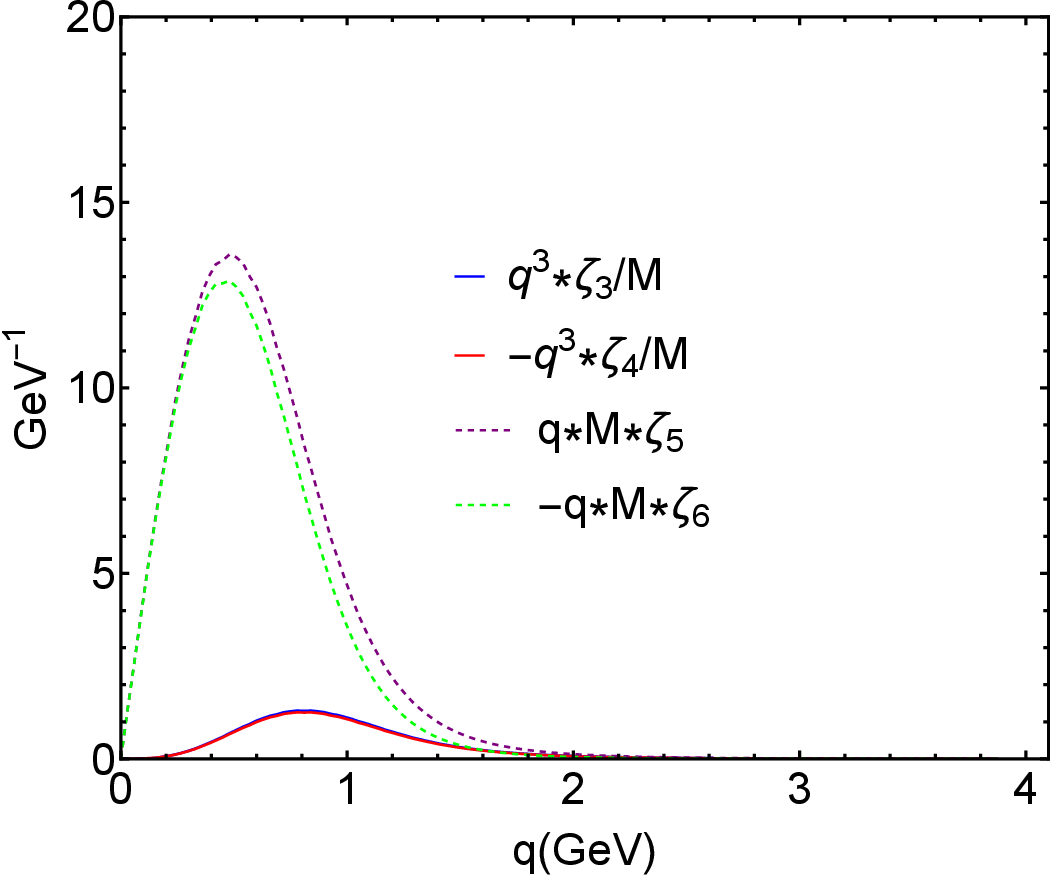}
    \includegraphics[width=5.3cm]{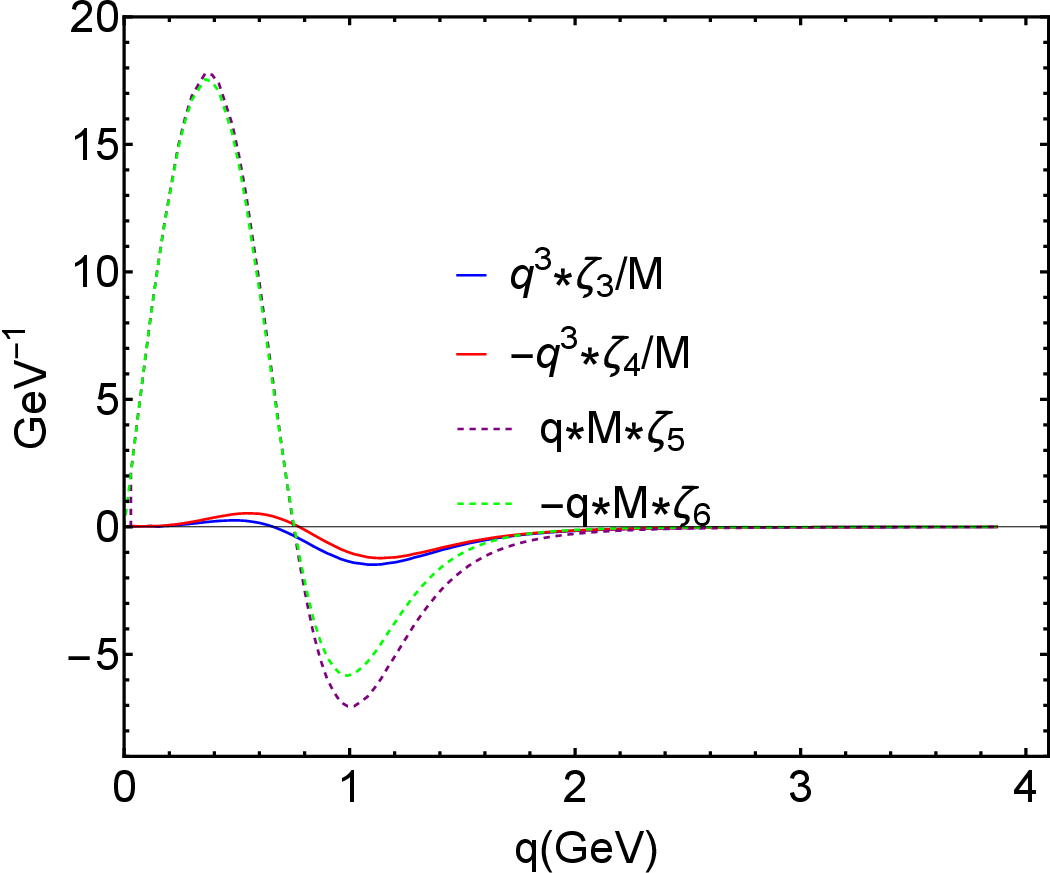}
    \includegraphics[width=5.3cm]{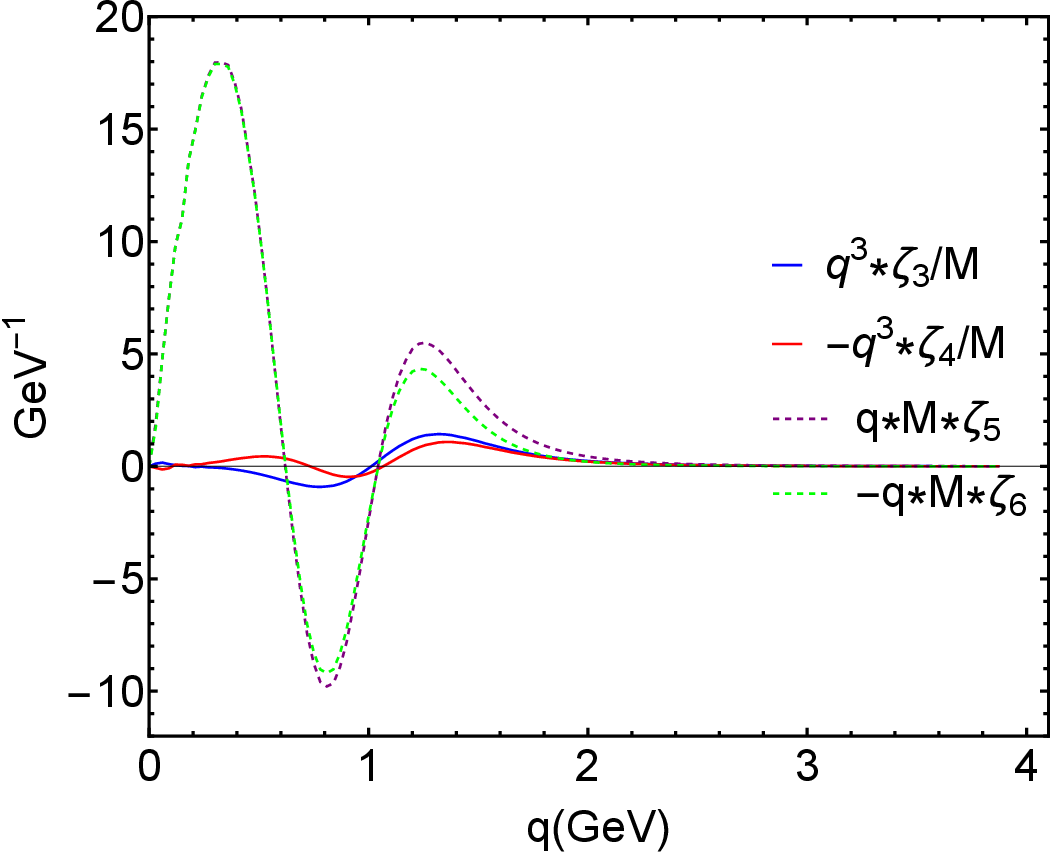}
    \centering \caption{The four independent radial wave functions for the $2^+$ tensors $D_{s2}^{\star -}(1P)$  (left), $D_{s2}^{\star -}(2P)$ (middle) and the $D_{s2}^{\star -}(3P)$ (right).} \label{Ds2}
\end{figure}

For the $2^+$ tensors, in our {semi-relativistic method}, their wave functions contain 8 terms, of which 4 are independent. The four independent radial wave functions for $2^+$ mesons $\bar D_2^{\star 0}(nP)$ and $D_{s2}^{\star -}(nP)$ $(n=1,2,3)$ are shown in Figure \ref{D20} and Figure \ref{Ds2}, respectively.


Among the eight terms of the wave function for a tensor, $\zeta _5$ and $\zeta _6$ terms are $P$ waves, $\zeta _3$ and $\zeta _4$ terms are mixture of $P$ and $F$ waves, and others, $\zeta _1$, $\zeta _2$, $\zeta _7$ and $\zeta _8$ are $D$ waves. Figures \ref{D20} and \ref{Ds2} roughly show us that the $2^+$ wave function is dominated by $P$ wave, which is consistent with the description of a non-relativistic method, where only $P$ wave exists with $\zeta _5=-\zeta _6$. In order to show the proportion of different waves, we use the normalization formulas, Eq.(\ref{2++ norm}), Eq.(\ref{2++P norm}) and Eq.(\ref{2++F norm}), to calculate the their ratios, and the results are shown in Table \ref{2++ partial wave}. We can see that, the $P$ wave is dominant, the $D$ wave is also sizable, and the $F$ wave is very small.
\begin{table}[ht]
\centering \caption{Ratios between the partial waves in the $2^{+}$ wave function} \label{2++ partial wave}
\setlength{\tabcolsep}{6pt}
\renewcommand{\arraystretch}{1}
\begin{tabular}{|c|c|c|c|c|c|}
\hline
   $2^{+}$           &           &   $1P$                 &    $2P$              &     $3P$   \\ \hline
 $D_2^{\star 0}$     &  $P:D:F$  & 1~:~0.393~:~0.0729     &  1~:~0.461~:~0.0787  &  1~:~0.560~:~0.0640  \\ \hline
 $D_2^ {\star -}$    &  $P:D:F$  & 1~:~0.389~:~0.0732     &  1~:~0.456~:~0.0793  &  1~:~0.550~:~0.0657  \\ \hline
 $D_{s2}^{\star -}$  &  $P:D:F$  & 1~:~0.298~:~0.0743     &  1~:~0.349~:~0.0843  &  1~:~0.404~:~0.0813  \\ \hline
 $\chi_{c2}$         &  $P:D:F$  & 1~:~0.140~:~0.0551     &  1~:~0.160~:~0.0673  &  1~:~0.177~:~0.0726  \\ \hline
\end{tabular}
\end{table}

In the non-relativistic limit, only $P$ wave exists, our results confirm that the $P$ wave is dominant, so these states are marked as $1P$, $2P$ and $3P$ states in Figures \ref{D20}, \ref{Ds2} and in Tables \ref{mass}, \ref{2++ partial wave}, respectively. {Compared with the non-relativistic $P$ waves from $\zeta_5$ and $\zeta_6$ terms, the $D$ and $F$ waves, as well as the $P$ waves from $\zeta_3$ and $\zeta_4$ terms} in the $2^+$ wave function provide the relativistic correction. From Table \ref{2++ partial wave}, it can be seen that for the $1P$, $2P$, and $3P$ states, the proportion of $F$ wave is very small and can be ignored when precise calculation is not required.
Based on the proportions of $D$ wave in Table \ref{2++ partial wave}, we conclude that, the relativistic correction in $D_2^{\star}$ is a little larger than that of $D_{s2}^{\star}$, and much larger than that in $\chi_{c2}$. It also shows that the relativistic correction of the highly excited state is larger than that of the lowly excited one, and the latter is larger than that of the ground state.

\subsection{The branching ratios of the semileptonic decays}

With the numerical values of wave functions and the formula for the transition matrix element, Eq.(\ref{transi}), the calculation of the semileptonic decay is straightforward. We present our results for branching ratios and other theoretical predictions in Table \ref{All result}. It is observed that there are few theoretical results in the literature regarding the case of highly excited tensor particles as the final state. Almost all existing results focus on studying the $1P$ final state process, with significant differences in predictions from different models, especially for $B^+ \rightarrow \bar D_2^{\star 0}(1P)\ell \nu_{\ell}$, whose branching ratios vary from 1.01 to 38.0.
\begin{table}[ht]
\begin{center}
\caption{Branching ratios ($10^{-3}$) of semileptonic decays and ratios $\mathcal{R}({D}_{2}^{\star })$, $\mathcal{R}({D}_{s2}^{\star})$, and $\mathcal{R}({\chi}_{c2})$.}\label{All result}
\setlength{\tabcolsep}{3mm}{
\begin{tabular}{c c c c c c c}
\hline
\textbf{Process}& & & & & &\textbf{Ours}\\
\hline
$B^+ \rightarrow \bar D_2^{\star 0}(1P)\ell \nu_{\ell}$ & 4.5$\sim$8.0\cite{Morenas} & $38.0$ \cite{Aliev}  & $1.01$ \cite{sundu} & $12.3$ \cite{cll}a   & $6.3$ \cite{cll}b    & $2.99$ \\
$B^+ \rightarrow\bar D_2^{\star 0}(1P)\tau \nu_{\tau}$ &  &$1.5$ \cite{Aliev}  & $0.16$ \cite{sundu} & $0.49$ \cite{cll}a & $0.22$ \cite{cll}b & $0.135$\\
$\frac{\mathcal{B}\left(B^{+} \rightarrow \bar{D}_{2}^{\star 0}(1P)\tau \nu_{\tau}\right)}{\mathcal{B}\left(B^{+} \rightarrow \bar{D}_{2}^{\star 0}(1P) \ell\nu_{\ell}\right)}$& &0.041\cite{Aliev}&0.16 \cite{sundu}&0.040\cite{cll}a&0.035\cite{cll}b&0.045
\\
$B^+ \rightarrow \bar D_2^{\star 0}(2P)\ell \nu_{\ell}$ &  &  &  & & & $0.075$ \\
$B^+ \rightarrow \bar D_2^{\star 0}(3P)\ell \nu_{\ell}$ &  &  &  & & & $0.0024$ \\ \hline
$B^0 \rightarrow D_2^{\star -}(1P)\ell \nu_{\ell}$ & 3.1(3.8)\cite{oda} & 2.5 \cite{D.Scora}  & 5.9 \cite{Ebert2} &5.86 \cite{Dong}  &  & $2.77$ \\
$B^0 \rightarrow D_2^{\star -}(1P)\tau \nu_{\tau}$ &  &  &  & & & $0.125$\\
$\frac{\mathcal{B}\left(B^{0} \rightarrow {D}_{2}^{\star -}(1P)\tau \nu_{\tau}\right)}{\mathcal{B}\left(B^{0} \rightarrow {D}_{2}^{\star -}(1P) \ell\nu_{\ell}\right)}$&  & & & & & $0.045$
\\
$B^0 \rightarrow D_2^{\star -}(2P)\ell \nu_{\ell}$ &  &  &  &  & & $0.070$ \\
$B^0 \rightarrow D_2^{\star -}(3P)\ell \nu_{\ell}$ &  &  &  &  & & $0.0022$ \\  \hline
$B_s^0 \rightarrow D_{s2}^{\star -}(1P)\ell \nu_{\ell}$ & 3.5 \cite{D.Scora} & $4.32$ \cite{Barakat:2022lmr} & $2.2$ \cite{Azizi:2014nta}  &6.7 \cite{Faustov}   &3.76 \cite{Segovia} & $3.76$ \\
$B_s^0 \rightarrow D_{s2}^{\star -}(1P)\tau \nu_{\tau}$ &   & 0.31 \cite{Barakat:2022lmr} & $0.926$\cite{Azizi:2014nta} &0.29 \cite{Faustov} & & $0.182$\\
$\frac{\mathcal{B}\left(B_s^{0} \rightarrow {D}_{s2}^{\star -}(1P)\tau \nu_{\tau}\right)}{\mathcal{B}\left(B_s^{0} \rightarrow {D}_{s2}^{\star -}(1P) \ell\nu_{\ell}\right)}$&  &0.071\cite{Barakat:2022lmr}&$0.42$ \cite{Azizi:2014nta}&0.043\cite{Faustov}&&0.048
\\
$B_s^0 \rightarrow D_{s2}^{\star -}(2P)\ell \nu_{\ell}$ & &  &  & & & $0.124$ \\
$B_s^0 \rightarrow D_{s2}^{\star -}(3P)\ell \nu_{\ell}$ & &  &  & & & $0.0047$ \\ \hline
$B_c^+ \rightarrow \chi_{c2}(1P)\ell \nu_{\ell}$& 1.0 \cite{D.Scora} &  1.6 \cite{Ebert}  & 1.7 \cite{Ivanov} & 1.3 \cite{ Hernandez} & 2.12 \cite{changwang2}& $1.82$  \\
$B_c^+ \rightarrow \chi_{c2}(1P)\tau \nu_{\tau}$&  &  0.093\cite{Ebert}  & 0.082\cite{Ivanov} & 0.093\cite{ Hernandez} & 0.33 \cite{changwang2}& $0.108$  \\
$\frac{\mathcal{B}\left(B_c^{+} \rightarrow \chi_{c2}(1P)\tau \nu_{\tau}\right)}{\mathcal{B}\left(B_c^{+} \rightarrow \chi_{c2}(1P) \ell\nu_{\ell}\right)}$&  &0.058\cite{Ebert}&0.048\cite{Ivanov}&0.072\cite{ Hernandez}&0.15 \cite{changwang2}&0.059
\\
$B_c^+ \rightarrow \chi_{c2}(2P)\ell \nu_{\ell}$& & 0.033\cite{Ebert} &  & & & $0.187$  \\
$B_c^+ \rightarrow \chi_{c2}(3P)\ell \nu_{\ell}$& &  &  & & & $0.0271$  \\
\hline
\end{tabular}}
\end{center}
\end{table}

Currently, only the production of the ground state $D_2^{\star}(1P)$ in the semileptonic decay of the $B$ meson and its cascade strong decay have been detected in experiments. The decay chains are $B \rightarrow D_2^{\star }(2460) \ell \nu_\ell$, $D_2^{\star }(2460) \rightarrow D \pi $. The averaged experimental results are \cite{PDG}
\begin{eqnarray}
&&\mathcal{B}\left(B^{+} \rightarrow \bar{D}_{2}^{\star 0} \ell^{+} \nu_{\ell}\right) \mathcal{B}\left(\bar{D}_{2}^{\star 0} \rightarrow D^{-} \pi^{+}\right) =  (1.53 \pm 0.16) \times 10^{-3}, \nonumber\\
&&\mathcal{B}\left(B^{+} \rightarrow \bar{D}_{2}^{\star 0} \ell^{+} \nu_{\ell}\right) \mathcal{B}\left(\bar{D}_{2}^{\star 0} \rightarrow D^{\star-} \pi^{+}\right) =  (1.01 \pm 0.24) \times 10^{-3}, \\
&&\mathcal{B}\left(B^{0} \rightarrow D_{2}^{\star-} \ell^{+} \nu_{\ell}\right) \mathcal{B}\left(D_{2}^{\star-} \rightarrow \bar{D}^{0} \pi^{-}\right)  = (1.21 \pm 0.33) \times 10^{-3}, \nonumber\\
&&\mathcal{B}\left(B^{0} \rightarrow D_{2}^{\star-} \ell^{+} \nu_{\ell}\right) \mathcal{B}\left(D_{2}^{\star-} \rightarrow \bar{D}^{\star 0} \pi^{-}\right) = (0.68 \pm 0.12) \times 10^{-3}.
\end{eqnarray}

The mass of $D_2^{\star}$ is above the thresholds of $D \pi$ and $D^\star \pi$, so $D_2^{\star}$ has the OZI-allowed strong decay channels $D_2^{\star} \rightarrow D \pi$ and $D_2^{\star} \rightarrow D^\star \pi$, which are the dominant decay processes of $D_2^{\star}$. Ref. \cite{sicheng zhang} predicted {$\mathcal{B}\left(\bar{D}_{2}^{\star 0} \rightarrow D^{-} \pi^{+}\right)=\mathcal{B}\left(D_{2}^{\star-} \rightarrow \bar{D}^{0} \pi^{-}\right)=44.5\%$ and $\mathcal{B}\left(\bar{D}_{2}^{\star 0} \rightarrow D^{\star-} \pi^{+}\right)=\mathcal{B}\left(D_{2}^{\star-} \rightarrow \bar{D}^{\star 0} \pi^{-}\right)=21.0\%$.} Using these, our theoretical predictions are
\begin{eqnarray}
&&\mathcal{B}\left(B^{+} \rightarrow \bar{D}_{2}^{\star 0} \ell^{+} \nu_{\ell}\right) \mathcal{B}\left(\bar{D}_{2}^{\star 0} \rightarrow D^{-} \pi^{+}\right) =  1.33 \times 10^{-3}, \nonumber\\
&&\mathcal{B}\left(B^{+} \rightarrow \bar{D}_{2}^{\star 0} \ell^{+} \nu_{\ell}\right) \mathcal{B}\left(\bar{D}_{2}^{\star 0} \rightarrow D^{\star-} \pi^{+}\right) =  0.628 \times 10^{-3}, \\
&&\mathcal{B}\left(B^{0} \rightarrow D_{2}^{\star-} \ell^{+} \nu_{\ell}\right) \mathcal{B}\left(D_{2}^{\star-} \rightarrow \bar{D}^{0} \pi^{-}\right)  = 1.23 \times 10^{-3}, \nonumber\\
&&\mathcal{B}\left(B^{0} \rightarrow D_{2}^{\star-} \ell^{+} \nu_{\ell}\right) \mathcal{B}\left(D_{2}^{\star-} \rightarrow \bar{D}^{\star 0} \pi^{-}\right) = 0.582 \times 10^{-3}.
\end{eqnarray}

The first two are slightly smaller than the experimental values, while the last two are in good agreement with the experimental data.

Similarly, using $\mathcal{B}\left({D}_{s2}^{\star -} \rightarrow \bar {D}^{0} K^{-}\right)=48.7\%$ and $\mathcal{B}\left({D}_{s2}^{\star -} \rightarrow D^{-} \bar{K}^{0}\right)=44.1\%$ from Ref.~\cite{sicheng zhang}, we obtain
\begin{eqnarray}
&&\mathcal{B}\left(B_s^{0} \rightarrow D_{s2}^{\star-} \ell^{+} \nu_{\ell}\right) \mathcal{B}\left(D_{s2}^{\star-} \rightarrow \bar{D}^{0} K^{-}\right)  = 1.83 \times 10^{-3}, \nonumber\\
&&\mathcal{B}\left(B_s^{0} \rightarrow D_{s2}^{\star-} \ell^{+} \nu_{\ell}\right) \mathcal{B}\left(D_{s2}^{\star-} \rightarrow {D}^{-} \bar{K}^{0}\right) = 1.66 \times 10^{-3}.
\end{eqnarray}

Compared to the ground $1P$ final state case, our results show that the branching ratio of the process with a highly excited final state ($2P$ or $3P$) is very small. The small branching ratio may be caused by the node structures (see Figures \ref{D20} and \ref{Ds2}) in the wave functions of the excited $2P$ and $3P$ mesons. The contributions of the wave functions on both sides of the node cancel each other, resulting in a very small branching ratio.

In Table \ref{All result}, the ratios $\mathcal{R}(\bar{D}_{2}^{\star 0})$, $\mathcal{R}({D}_{2}^{\star -})$, $\mathcal{R}({D}_{s2}^{\star})$, and $\mathcal{R}({\chi}_{c2})$ are also listed, where, for example,
\begin{equation}
\mathcal{R}(\bar{D}_{2}^{\star 0})=\frac{\mathcal{B}\left(B^{+} \rightarrow \bar{D}_{2}^{\star 0}(1P)\tau \nu_{\tau}\right)}{\mathcal{B}\left(B^{+} \rightarrow \bar{D}_{2}^{\star 0}(1P) \ell\nu_{\ell}\right)}.
\end{equation}
The ratio $\mathcal{R}$ may cancel some model-dependent factors, which can be seen from the results of Refs.~\cite{Aliev}, \cite{cll}, and ours. The branching ratios are much different, but the $\mathcal{R}(\bar{D}_{2}^{\star 0})$ values is  around 0.04, which is very close to each other. We have similar conclusions for $\mathcal{R}({D}_{s2}^{\star})$ and $\mathcal{R}({\chi}_{c2})$.

We have pointed out that the relativistic corrections of excited states are greater than those of ground states \cite{part wave,wangv}. And there are still significant differences in semileptonic decays between theoretical results, especially for the $B$ decays. The differences may be caused by the relativistic corrections. So for these processes, which contains excited states, we need a more careful study, especially the relativistic corrections. In the following, we will study the detailed contributions of different partial waves.

\subsection{Contributions of different partial waves}

We provide the proportions of different partial waves in the wave function, allowing us to roughly estimate the magnitude of the relativistic correction. However, this does not represent the true relativistic correction of particles in interaction, as different partial waves behave differently in interactions. What we need is the overlapping integration between wave functions, not the individual wave functions themselves. Therefore, using some transition processes as examples, we present the detailed contributions of partial waves.

\subsubsection{$B^+ \rightarrow\bar D_2^{\star 0}(1P)\ell^{+} \nu_{\ell}$}

Table \ref{0- partial wave} shows that the wave function of $B^+$ is dominanted by $S$ wave ($A_1$ and $A_2$ terms) but mixed with $P$ wave ($A_3$ and $A_4$ terms), their ratio is $S : P = 1 : 0.339$. For $\bar D_2^{\star 0}$,  $P$ wave is dominant and mixed with $D$ and $F$ waves, $P : D : F = 1:0.393:0.0729 $.

To see the detail of the transition $B^+ \rightarrow \bar D_2^{\star 0}(1P)$, we will carefully study the overlapping integral of $(S+P)\times(P'+D'+F')$, where to distinguish between the initial and final states, we use `prime' to represent the final state. We show some of the detailed contributions of different partial waves to the branching ratio of $B^+ \rightarrow \bar D_2^{\star 0}(1P)\ell^{+} \nu_{\ell}$ in Table \ref{B+ to 1P}. Where `whole' means the complete wave function, while `$S$ wave' in the column or `$P'$ wave' in the row means the corresponding result is obtained only using the $S$ or $P'$ wave and ignoring others, etc.
\begin{table}[ht]
\centering
\caption{Contributions of partial waves to the branching ratio of $B^+ \rightarrow \bar D_2^{\star 0}(1P)\ell^{+} \nu_{\ell}$ (in $10^{-4}$).} \label{B+ to 1P}
\setlength{\tabcolsep}{6pt}
\renewcommand{\arraystretch}{1}
\begin{tabular}{|c|c|c|c|c|}
\hline
{\diagbox{$0^-$}{$2^+$}} & whole & $P'$ wave & $D'$ wave ($B_1$,$B_2$,$B_7$,$B_8$) & $F'$ wave \\ \hline
whole & 29.9 & 15.1 & 3.06 & 0.063 \\ \hline
$S$ wave ($A_1$,$A_2$) & 18.9 & 35.6 & 3.91 & 0.0026 \\ \hline
$P$ wave ($A_3$,$A_4$) & 1.48 & 4.45 & 10.1 & 0.048 \\ \hline
\end{tabular}
\end{table}

From Table \ref{B+ to 1P}, we can see that the dominant $S$ partial wave in the $B^+$ state and $P'$ wave in $\bar D_2^{\star 0}(1P)$ provide the dominant contribution. The $P$ wave in $B^+$ and $D'$ wave in $\bar D_2^{\star 0}(1P)$ give the main relativistic corrections, while the $F'$ partial wave in $\bar D_2^{\star 0}(1P)$ has a tiny contribution, which can be safely ignored.

{For the $B^+$ meson, its non-relativistic wave function only contains $S$ wave, so its $P$ wave provides the relativistic correction. For the $D_2^{\star 0}(1P)$, the situation is relatively complex. In the non-relativistic case, its wave function only contains $P'$ waves, but only $P'$ waves from $\zeta_5$ and $\zeta_6$ terms, without the ones from $\zeta_3$ and $\zeta_4$ terms, see the formula in Eq.(\ref{2++P}). Therefore, in the non-relativistic scenario, the branching ratio of $S\times P'$ changes from $35.6 \times10^{-4}$  in Table \ref{B+ to 1P} to $37.2 \times10^{-4}$. Our complete branching ratio is $\mathcal{B}_{\text{rel}}=29.9 \times10^{-4}$,} so the relativistic effect can be calculated as
\begin{equation}\label{B+ to 1P rel}
\frac{\mathcal{B}_{\text{non-rel}}-\mathcal{B}_{\text{rel}}}{\mathcal{B}_{\text{rel}}} = 24.4\%,
\end{equation}
which is significant but not as large as expected. This might be due to two possible reasons. First, there could be a cancellation between relativistic corrections; for example, the contribution of ovarlapping $P\times P'$ is $4.45\times10^{-4}$, $P\times D'$ is $10.1\times10^{-4}$, while their sum contribution $P\times (P'+D')$ is $1.48\times10^{-4}$. Second, from Table \ref{B+ to 1P}, we can see that the main relativistic correction is from the interaction $P\times D'$, not from $S\times D'$ or $P\times P'$.

\subsubsection{$B^+ \rightarrow\bar D_2^{\star 0}(2P)\ell^{+} \nu_{\ell}$}

Table \ref{B+ to 2P} shows the details of the decay $B^+ \rightarrow \bar D_2^{\star 0}(2P)\ell^{+} \nu_{\ell}$. Compared with the case of $\bar D_2^{\star 0}(1P)$ final state, the contributions of all the partial waves are much smaller. The main reason is that there are nodes in all the partial wave functions of the $2P$ state, and the contributions of the wave functions before and after the nodes cancel each other, resulting in a very small branching ratio. In addition, the mass of $\bar D_2^{\star 0}(2P)$ is heavier than that of $\bar D_2^{\star 0}(1P)$, and the phase space of the decay $B^+ \rightarrow \bar D_2^{\star 0}(2P)\ell^{+} \nu_{\ell}$ is smaller than that of $B^+ \rightarrow \bar D_2^{\star 0}(1P)\ell^{+} \nu_{\ell}$.
\begin{table}[ht]
\centering
\caption{Contributions of partial waves to the branching ratio of $B^+ \rightarrow \bar D_2^{\star 0}(2P)\ell^{+} \nu_{\ell}$ (in $10^{-4}$).} \label{B+ to 2P}
\setlength{\tabcolsep}{6pt}
\renewcommand{\arraystretch}{1}
\begin{tabular}{|c|c|c|c|c|}
\hline
{\diagbox{$0^-$}{$2^+$}} & whole & $P'$ wave & $D'$ wave ($B_1$,$B_2$,$B_7$,$B_8$) & $F'$ wave \\ \hline
whole & 0.752 & 0.0087 & 0.573 & 0.0033 \\ \hline
$S$ wave ($A_1$,$A_2$) & 0.015 & 0.0357 & 0.006 & 0.0001 \\ \hline
$P$ wave ($A_3$,$A_4$) & 0.557 & 0.0106 & 0.691 & 0.0026 \\ \hline
\end{tabular}
\end{table}

It can be seen from Table \ref{B+ to 2P} that the largest contribution does not come from the non-relativistic term $S\times P'$, nor from the relativistic corrections $S\times D'$ and $P\times P'$, but from the relativistic correction $P\times D'$.
The results show that the node structure has a more severe inhibitory effect on $S\times P'$ than on $P\times D'$, leading to the latter providing the maximum contribution and a large relativistic effect in this process.
%\begin{equation}\label{B+ to 2P rel}
% \frac{\mathcal{B}_{\text{rel}}-\mathcal{B}_{\text{non-rel}}}{\mathcal{B}_{\text{rel}}} = 92.3 \%.
%\end{equation}

\subsubsection{$B_s^0 \rightarrow D_{s2}^{\star -}(1P)\ell^{+} \nu_{\ell}$}

\begin{table}[ht]
\centering
\caption{Contributions of partial waves to the branching ratio of $B_s^0 \rightarrow D_{s2}^{\star -}(1P)\ell^{+} \nu_{\ell}$ (in $10^{-4}$)}\label{Bs to 1P}
\setlength{\tabcolsep}{6pt}
\renewcommand{\arraystretch}{1}
\begin{tabular}{|c|c|c|c|c|}
\hline
{\diagbox{$0^-$}{$2^+$}} & whole & $P'$ wave & $D'$ wave ($B_1$,$B_2$,$B_7$,$B_8$) & $F'$ wave \\ \hline
whole & 37.6 & 29.7 & 1.23 & 0.0359 \\ \hline
$S$ wave ($A_1$,$A_2$) & 29.0 & 48.0 & 2.90 & 0.0168 \\ \hline
$P$ wave ($A_3$,$A_4$) & 0.620 & 2.29 & 4.80 & 0.0309 \\ \hline
\end{tabular}
\end{table}

Table \ref{Bs to 1P} shows that, similar to the process of $B^+ \rightarrow \bar D_2^{\star 0}(1P)\ell^{+} \nu_{\ell}$, the overlap of $S\times P'$ provides the dominant contribution to $B_s^0 \rightarrow D_{s2}^{\star -}(1P)\ell^{+} \nu_{\ell}$, which is mainly non-relativistic. All other contributions are relativistic corrections, with $P\times D'$ being the largest, followed by $S\times D'$ and $P\times P'$, while the contribution of $F'$ wave can be safely ignored.
{ In the non-relativistic limit, the branching ratio of $S\times P'$ should be changed from $48.0 \times10^{-4}$ in Table \ref{Bs to 1P} to $48.4\times10^{-4}$, so} the relativistic  effect is
\begin{equation}\label{Bs to 1P rel}
 \frac{\mathcal{B}_{\text{non-rel}}-\mathcal{B}_{\text{rel}}}{\mathcal{B}_{\text{rel}}} = 28.8\%,
\end{equation}
which is also not as large as we expected, but a little larger than those of $B^+ \rightarrow \bar D_2^{\star 0}(1P)$. However, we cannot simply conclude that the relativistic  effect of the former is greater than that of the latter. When we look at the details of relativistic  corrections, compared with the non-relativistic contribution, the contributions of $P\times D'$, $S\times D'$, and $P\times P'$ in the process $B_s^0 \rightarrow D_{s2}^{\star -}(1P)\ell^{+} \nu_{\ell}$ are much smaller than those in $B^+ \rightarrow \bar D_2^{\star 0}(1P)\ell^{+} \nu_{\ell}$, respectively. However, when summing them up, the overall result of the latter is smaller, due to cancellation.

\subsubsection{$B_s^0 \rightarrow D_{s2}^{\star -}(2P)\ell^{+} \nu_{\ell}$}

\begin{table}[ht]
\centering
\caption{Contributions of partial waves to the branching ratio of $B_s^0 \rightarrow D_{s2}^{\star -}(2P)\ell^{+} \nu_{\ell}$ (in $10^{-4}$)}\label{Bs to 2P}
\setlength{\tabcolsep}{6pt}
\renewcommand{\arraystretch}{1}
\begin{tabular}{|c|c|c|c|c|}
\hline
{\diagbox{$0^-$}{$2^+$}} & whole & $P'$ wave & $D'$ wave ($B_1$,$B_2$,$B_7$,$B_8$) & $F'$ wave \\ \hline
whole & 1.24 & 0.137 & 0.546 & 0.00411 \\ \hline
$S$ wave ($A_1$,$A_2$) & 0.154 & 0.327 & 0.0241 & 0.00105 \\ \hline
$P$ wave ($A_3$,$A_4$) & 0.528 & 0.0414 & 0.789 & 0.00545 \\ \hline
\end{tabular}
\end{table}

Table \ref{Bs to 2P} illustrates the scenario of $B_s^0 \rightarrow D_{s2}^{\star -}(2P)\ell^{+} \nu_{\ell}$, which bears resemblance to the case of $B^+ \rightarrow \bar D_2^{star 0}(2P)\ell^{+} \nu_{\ell}$. Notably, significant relativistic  effects are observed. The primary contributions to the branching ratio arise from relativistic  corrections, particularly the $P\times D'$ term, rather than non-relativistic contributions.
%\begin{equation}\label{Bs to 2P rel}
% \frac{\mathcal{B}_{\text{rel}}-\mathcal{B}_{\text{non-rel}}}{\mathcal{B}_{\text{rel}}} = 71.3 \%,
%\end{equation}

\subsubsection{$B_c^+ \rightarrow \chi_{c2}(1P)\ell^{+} \nu_{\ell}$}

\begin{table}[ht]
\centering
\caption{Contributions of partial waves to the branching ratio of $B_c^+ \rightarrow \chi_{c2}(1P)\ell^{+} \nu_{\ell}$ (in $10^{-4}$).}\label{Bc to 1P}
\setlength{\tabcolsep}{6pt}
\renewcommand{\arraystretch}{1}
\begin{tabular}{|c|c|c|c|c|}
\hline
{\diagbox{$0^-$}{$2^+$}} & whole & $P'$ wave & $D'$ wave ($B_1$,$B_2$,$B_7$,$B_8$) & $F'$ wave \\ \hline
whole & 18.2 & 19.1 & 0.110 & 0.00246 \\ \hline
$S$ wave ($A_1$,$A_2$) & 18.3 & 21.9 & 0.255 & 0.00012 \\ \hline
$P$ wave ($A_3$,$A_4$) & 0.0211 & 0.108 & 0.0761 & 0.00181 \\ \hline
\end{tabular}
\end{table}

{In contrast to the value $21.9 \times10^{-4}$ of $S\times P'$ shown in Table \ref{Bc to 1P}, in the non-relativistic limit, the branching ratio for $S\times P'$ is $22.2\times10^{-4}$,} so we obtain
\begin{equation}
 \frac{\mathcal{B}_{\text{non-rel}}-\mathcal{B}_{\text{rel}}}{\mathcal{B}_{\text{rel}}} = 22.1\%
\end{equation}
for $B_c^+ \rightarrow \chi_{c2}(1P)\ell^{+} \nu_{\ell}$. This value seems not much different from that of $B^+ \rightarrow \bar D_2^{\star 0}(1P)\ell^{+} \nu_{\ell}$ or $B_s^0 \rightarrow D_{s2}^{\star -}(1P)\ell^{+} \nu_{\ell}$. However, from Tables \ref{B+ to 1P}, \ref{Bs to 1P}, and \ref{Bc to 1P}, we can see that, although the complete branching ratios and non-relativistic results do not differ significantly, each relativistic  correction in $B_c^+ \rightarrow \chi_{c2}(1P)\ell^{+} \nu_{\ell}$ is much smaller than that in $B^+ \rightarrow \bar D_2^{\star 0}(1P)\ell^{+} \nu_{\ell}$ or in $B_s^0 \rightarrow D_{s2}^{\star -}(1P)\ell^{+} \nu_{\ell}$, respectively. We also note that the largest relativistic  correction comes from $S\times D'$, not $P\times D'$.

\subsubsection{$B_c^+ \rightarrow \chi_{c2}(2P)\ell^{+} \nu_{\ell}$}

\begin{table}[ht]
\centering
\caption{Contributions of partial waves to the branching ratio of $B_c^+ \rightarrow \chi_{c2}(2P)\ell^{+} \nu_{\ell}$ (in $10^{-4}$).}\label{Bc to 2P}
\setlength{\tabcolsep}{6pt}
\renewcommand{\arraystretch}{1}
\begin{tabular}{|c|c|c|c|c|}
\hline
{\diagbox{$0^-$}{$2^+$}} & whole & $P'$ wave & $D'$ wave ($B_1$,$B_2$,$B_7$,$B_8$) & $F'$ wave \\ \hline
whole & 1.87 & 1.50 & 0.0245 & 0.00101 \\ \hline
$S$ wave ($A_1$,$A_2$) & 1.54 & 1.79 & 0.0241 & 0.00002 \\ \hline
$P$ wave ($A_3$,$A_4$) & 0.0169 & 0.0126 & 0.0523 & 0.00085 \\ \hline
\end{tabular}
\end{table}

From Table \ref{Bc to 2P}, we can see that, unlike the cases of $B^+ \rightarrow \bar D_2^{\star 0}(2P)\ell^{+} \nu_{\ell}$ and $B_s^0 \rightarrow D_{s2}^{\star -}(2P)\ell^{+} \nu_{\ell}$, the non-relativistic $S\times P'$ in $B_c^+ \rightarrow \chi_{c2}(2P)\ell^{+} \nu_{\ell}$ still contributes the most, much larger than the relativistic  corrections, indicating that the node structure has different effects on the processes $B_c^+ \rightarrow \chi_{c2}(2P)\ell^{+} \nu_{\ell}$ and $B^+ \rightarrow \bar D_2^{\star 0}(2P)\ell^{+} \nu_{\ell}$ (or $B_s^0 \rightarrow D_{s2}^{\star -}(2P)\ell^{+} \nu_{\ell}$).

\subsection{CONCLUSION}

{We present a semi-relativistic study on the semileptonic decays of heavy pseudoscalars $B^+$, $B^0$, $B_s^0$, and $B_c^+$ to $1P$, $2P$, and $3P$ $2^+$ tensors caused by the transition of $\bar b\to \bar c$ by using the instantaneous Bethe$-$Salpeter method.} We obtain $\mathcal{B}(B^+ \rightarrow\bar D_2^{\star 0}(1P)\ell^{+} \nu_{\ell})=2.99\times 10^{-3}$ and  $\mathcal{B}(B^0 \rightarrow\bar D_2^{\star -}(1P)\ell^{+} \nu_{\ell})=2.77\times 10^{-3}$, which are in good agreement with the experimental data. For the undetected channels, our results are $\mathcal{B}\left(B_s^{0} \rightarrow D_{s2}^{\star-}(1P) \ell^{+} \nu_{\ell}\right) = 3.76\times 10^{-3}$ and $\mathcal{B}\left(B_c^+ \rightarrow \chi_{c2}(1P)\ell^{+} \nu_{\ell}\right) = 1.82\times 10^{-3}$. For the decays to the $2P$ and $3P$ states, all branching ratios are very small and cannot be detected in current experiments.

In this paper, we focus on studying the different partial waves in the {Salpeter wave functions} and their contributions in semileptonic decays.

(1) In the wave function for the $0^-$ states, $B^+$, $B^0$, $B_s^0$, or $B_c^+$, $S$-wave is dominant and provides the non-relativistic contribution; $P$-wave is sizable and gives the relativistic  correction. While for the $2^+$ states, $\bar D_2^{\star 0}$, $D_2^{\star -}$, $D_{s2}^{\star -}$, or $\chi_{c2}$, $P$-wave is dominant, combined with sizable $D$-wave, and tiny $F$-wave, where {$P$-wave from $\zeta_5$ and $\zeta_6$ terms} gives the non-relativistic contribution, others contribute to the relativistic  corrections.

(2) We note that, considering only the wave functions, the relativistic  corrections for $B$, $B_s$, $\bar D_2^{\star}$, and $D_{s2}^{\star}$ mesons are large, while the relativistic  corrections for $\chi_{c2}$ and $B_c$ are small. However, when calculating the transition process, the overlapping integration of wave functions plays a major role. Thus, we obtain similar relativistic  effects, for example, $24.4\%$ for $B^+ \rightarrow\bar D_2^{\star 0}(1P)\ell^{+} \nu_{\ell}$, $28.8\%$ for $B_s^{0} \rightarrow D_{s2}^{\star-}(1P) \ell^{+} \nu_{\ell}$ and $22.1\%$ for $B_c^+ \rightarrow \chi_{c2}(1P)\ell^{+} \nu_{\ell}$.

(3) When we look at the details, there are significant differences. For example, in the transition of $B\to D_2^{\star}(1P)$, the individual contributions of relativistic partial waves are significant, while in the overall result, they are in a cancelling relationship, resulting in a small overall relativistic  effect. While in $B_c^+ \rightarrow \chi_{c2}(1P)$, the individual contributions of relativistic waves are small, directly leading to a small overall relativistic  effect.

(4) When the process contains a radially excited state, the node structure in the wave function of the radially excited state plays an overwhelming role, resulting in a very small branching ratio.

\vspace{0.7cm} {\bf Acknowledgments}

This work was supported in part by the National Natural Science Foundation of China (NSFC) under the Grants Nos. 12075073, 12375085 and 11865001, the Natural Science Foundation of Hebei province under the Grant No. A2021201009.

\appendix
\section{Salpeter equations}

\begin{align}
MF_{1}(q_{_{\perp}} ) & = (\omega_{1} + \omega_{2} )F_{1}(q_{_{\perp}} ) + \int\frac{d\vec{k}}{(2\pi)^{3}}\frac{1}{24\omega_{1}\omega_{2}}\bigg\{ 4{(V_S-V_V)}(e_{1}m_{2} + e_{2}m_{1} )\bigg[ - (F_{3}(k_{_{\perp}} ) - F_{4}(k_{_{\perp}} ) )\nonumber \\
 & \quad - \frac{m_{1} - m_{2}}{e_{1} + e_{2}}(F_{1}(k_{_{\perp}} ) + F_{2}(k_{_{\perp}} ) ) {-}(\frac{e_{1} - e_{2}}{m_{1} + m_{2}}(F_{1}(k_{_{\perp}} ) - F_{2}(k_{_{\perp}} ) ) + (F_{3}(k_{_{\perp}} ) + F_{4}(k_{_{\perp}} ) ))\nonumber \\
 & \quad \times \frac{\omega_{1} + \omega_{2}}{e_{1} + e_{2}}\bigg]\frac{q_{_{\perp}}^{2}\vec{k} \cdot \vec{q}}{k_{_{\perp}}^{4}} {-}9{(V_S+V_V)}\bigg[(F_{1}(k_{_{\perp}} ) + F_{2}(k_{_{\perp}} ) )(q_{_{\perp}}^{2} + m_{1}m_{2} - \omega_{1}\omega_{2} )\nonumber \\
 & \quad + (F_{1}(k_{_{\perp}} ) - F_{2}(k_{_{\perp}} ) )(\omega_{1}m_{2} - \omega_{2}m_{1} )\frac{e_{1} - e_{2}}{m_{1} + m_{2}}\bigg](\frac{(\vec{k} \cdot \vec{q} )^{2}}{k_{_{\perp}}^{4}} - \frac{q_{_{\perp}}^{2}}{3k_{_{\perp}}^{2}} )+3{(V_S-V_V)}\nonumber \\
 & \quad \times (e_{1}m_{2} - e_{2}m_{1} )\bigg[(F_{1}(k_{_{\perp}} ) + F_{2}(k_{_{\perp}} ) )\frac{5m_{1} + m_{2}}{e_{1} + e_{2}} + 2(F_{3}(k_{_{\perp}} ) - F_{4}(k_{_{\perp}} ) ) + (\frac{5e_{1} + e_{2}}{m_{1} + m_{2}}\nonumber \\
 & \quad \times (F_{1}(k_{_{\perp}} ) - F_{2}(k_{_{\perp}} ) ) + 2(F_{3}(k_{_{\perp}} ) + F_{4}(k_{_{\perp}} ) ))\frac{\omega_{1} + \omega_{2}}{e_{1} + e_{2}}\bigg](\frac{(\vec{k} \cdot \vec{q} )^{3}}{k_{_{\perp}}^{6}} - \frac{q_{_{\perp}}^{2}\vec{k} \cdot \vec{q}}{3k_{_{\perp}}^{4}} )\bigg\};
\end{align}

\begin{align}
MF_{2}(q_{_{\perp}} ) & = - (\omega_{1} + \omega_{2} )F_{2}(q_{_{\perp}} ) - \int\frac{d\vec{k}}{(2\pi)^{3}}\frac{1}{24\omega_{1}\omega_{2}}\bigg\{ 4{(V_S-V_V)}(e_{1}m_{2} + e_{2}m_{1} )\bigg[ - (F_{3}(k_{_{\perp}} ) - F_{4}(k_{_{\perp}} ) )\nonumber \\
 & \quad - \frac{m_{1} - m_{2}}{e_{1} + e_{2}}(F_{1}(k_{_{\perp}} ) + F_{2}(k_{_{\perp}} ) ){+} (\frac{e_{1} - e_{2}}{m_{1} + m_{2}}(F_{1}(k_{_{\perp}} ) - F_{2}(k_{_{\perp}} ) ))\nonumber \\
 & \quad \times \frac{\omega_{1} + \omega_{2}}{e_{1} + e_{2}}\bigg]\frac{q_{_{\perp}}^{2}\vec{k} \cdot \vec{q}}{k_{_{\perp}}^{4}} {-}9{(V_S+V_V)}\bigg[(F_{1}(k_{_{\perp}} ) + F_{2}(k_{_{\perp}} ) )(q_{_{\perp}}^{2} + m_{1}m_{2} - \omega_{1}\omega_{2} )\nonumber \\
 & \quad - (F_{1}(k_{_{\perp}} ) - F_{2}(k_{_{\perp}} ) )(\omega_{1}m_{2} - \omega_{2}m_{1} )\frac{e_{1} - e_{2}}{m_{1} + m_{2}}\bigg](\frac{(\vec{k} \cdot \vec{q} )^{2}}{k_{_{\perp}}^{4}} - \frac{q_{_{\perp}}^{2}}{3k_{_{\perp}}^{2}} ) +3{(V_S-V_V)}\nonumber \\
 & \quad \times (e_{1}m_{2} - 3e_{2}m_{1} )\bigg[(F_{1}(k_{_{\perp}} ) + F_{2}(k_{_{\perp}} ) )\frac{5m_{1} + m_{2}}{e_{1} + e_{2}} + 2(F_{3}(k_{_{\perp}} ) - F_{4}(k_{_{\perp}} ) )- (\frac{5e_{1} + e_{2}}{m_{1} + m_{2}}\nonumber \\
 & \quad \times (F_{1}(k_{_{\perp}} ) - F_{2}(k_{_{\perp}} ) ) + 2(F_{3}(k_{_{\perp}} ) + F_{4}(k_{_{\perp}} ) ))\frac{\omega_{1} + \omega_{2}}{e_{1} + e_{2}}\bigg](\frac{(\vec{k} \cdot \vec{q} )^{3}}{k_{_{\perp}}^{6}} - \frac{q_{_{\perp}}^{2}\vec{k} \cdot \vec{q}}{3k_{_{\perp}}^{4}} )\bigg\};
\end{align}

\begin{align}
M F_{3}(q_{_{\perp}} ) & = (\omega_{1} + \omega_{2} )F_{3}(q_{_{\perp}} ) + \int\frac{d\vec{k}}{(2\pi)^{3}}\frac{1}{24\omega_{1}\omega_{2}}\bigg\{{-}10 {(V_S+V_V)}\bigg[\bigg(\frac{m_{1} - m_{2}}{m_{1} + m_{2}}\frac{e_{1} - e_{2}}{e_{1} + e_{2}}\nonumber \\
 & \quad \times (F_{1}(k_{_{\perp}} ) + F_{2}(k_{_{\perp}} ) ) + \frac{e_{1} - e_{2}}{m_{1} + m_{2}}(F_{3}(k_{_{\perp}} ) - F_{4}(k_{_{\perp}} ) )\bigg)(m_{2}\omega_{1} - m_{1}\omega_{2} )\nonumber \\
 & \quad + \bigg(\frac{e_{1} - e_{2}}{m_{1} + m_{2}}(F_{1}(k_{_{\perp}} ) - F_{2}(k_{_{\perp}} ) ) + (F_{3}(k_{_{\perp}} ) + F_{4}(k_{_{\perp}} ) ) \bigg)(q_{_{\perp}}^{2} + m_{1}m_{2} - \omega_{1}\omega_{2} )\bigg]{\frac{(\vec{k} \cdot \vec{q})^2}{k_{_{\perp}}^{4}}}\nonumber \\
 & \quad {-}8{(V_S-V_V)}\frac{e_{2}m_{1} + e_{1}m_{2}}{(e_{1} + e_{2} )(m_{1} + m_{2} )}\bigg[ m_{2}((e_{1} - e_{2} )(F_{1}(k_{_{\perp}} ) - F_{2}(k_{_{\perp}} ) ) + (m_{1} + m_{2} )\nonumber \\
 & \quad \times (F_{3}(k_{_{\perp}} ) + F_{4}(k_{_{\perp}} ) ) ) + \omega_{2}((m_{1} - m_{2} )(F_{1}(k_{_{\perp}} ) + F_{2}(k_{_{\perp}} ) ) + (e_{1} + e_{2} )(F_{3}(k_{_{\perp}} ) - F_{4}(k_{_{\perp}} ) ) )\bigg]\nonumber \\
 & \quad \times \frac{q_{_{\perp}}^{2}\vec{k} \cdot \vec{q}}{k_{_{\perp}}^{4}} {+}10{(V_S-V_V)}(e_{2}m_{1} + e_{1}m_{2} )\bigg[\frac{e_{1} - e_{2}}{e_{1} + e_{2}}(F_{1}(k_{_{\perp}} ) - F_{2}(k_{_{\perp}} ) ) + \frac{m_{1} + m_{2}}{e_{1} + e_{2}}\nonumber \\
 & \quad \times (F_{3}(k_{_{\perp}} ) + F_{4}(k_{_{\perp}} ) ) + \frac{\omega_{1} + \omega_{2}}{m_{1} + m_{2}}(\frac{m_{1} - m_{2}}{e_{1} + e_{2}}(F_{1}(k_{_{\perp}} ) + F_{2}(k_{_{\perp}} ) ) + (F_{3}(k_{_{\perp}} ) - F_{4}(k_{_{\perp}} ) ) )\bigg]\nonumber \\
 & \quad \times \frac{q_{_{\perp}}^{2}\vec{k} \cdot \vec{q}}{k_{_{\perp}}^{4}} {-}2{(V_S-V_V)}(e_{2}m_{1} + e_{1}m_{2} )\bigg[\frac{5e_{1} + e_{2}}{e_{1} + e_{2}}(F_{1}(k_{_{\perp}} ) - F_{2}(k_{_{\perp}} ) ) + 2\frac{m_{1} + m_{2}}{e_{1} + e_{2}}\nonumber \\
 & \quad \times (F_{3}(k_{_{\perp}} ) + F_{4}(k_{_{\perp}} ) ) + \frac{\omega_{1} + \omega_{2}}{m_{1} + m_{2}}(\frac{5m_{1} + m_{2}}{e_{1} + e_{2}}(F_{1}(k_{_{\perp}} ) + F_{2}(k_{_{\perp}} ) ) + 2(F_{3}(k_{_{\perp}} ) - F_{4}(k_{_{\perp}} ) ) )\bigg]\nonumber \\
 & \quad \times \frac{q_{_{\perp}}^{2}\vec{k} \cdot \vec{q}}{k_{_{\perp}}^{4}} {+}3{(V_S+V_V)}\bigg[ (F_{1}(k_{_{\perp}} ) + F_{2}(k_{_{\perp}} ) )(m_{1}\omega_{2} - m_{2}\omega_{1} ) - \frac{e_{1} - e_{2}}{m_{1} + m_{2}}(F_{1}(k_{_{\perp}} ) - F_{2}(k_{_{\perp}} ) ) \nonumber  \\
 & \quad \times (- q_{_{\perp}}^{2} + m_{1}m_{2} - \omega_{1}\omega_{2} ) \bigg](\frac{q_{_{\perp}}^{2}}{k_{_{\perp}}^{2}} - \frac{3(\vec{k} \cdot \vec{q} )^{2}}{k_{_{\perp}}^{4}} ) +{(V_S+V_V)}\frac{1}{m_{1} + m_{2}}\bigg[\frac{e_{1} - e_{2}}{e_{1} + e_{2}}((m_{1} - m_{2} )\nonumber \\
 & \quad \times (F_{1}(k_{_{\perp}} ) + F_{2}(k_{_{\perp}} ) ) + (e_{1} + e_{2} )(F_{3}(k_{_{\perp}} ) - F_{4}(k_{_{\perp}} ) ) ) (m_{2}\omega_{1} - m_{1}\omega_{2} ) + ((e_{1} - e_{2} )\nonumber \\
 & \quad \times (F_{1}(k_{_{\perp}} ) - F_{2}(k_{_{\perp}} ) ) + (m_{1} + m_{2} )(F_{3}(k_{_{\perp}} ) + F_{4}(k_{_{\perp}} ) ) ) (q_{_{\perp}}^{2} + m_{1}m_{2} - \omega_{1}\omega_{2} )\bigg]\nonumber \\
 & \quad \times (\frac{3q_{_{\perp}}^{2}}{k_{_{\perp}}^{2}}{+} \frac{(\vec{k} \cdot \vec{q} )^{2}}{k_{_{\perp}}^{4}} ){+}2 {(V_S-V_V)}\frac{e_{2}m_{1} + e_{1}m_{2}}{(e_{1} + e_{2} )(m_{1} + m_{2} )}\bigg[ m_{2}((5e_{1} + e_{2} ) (F_{1}(k_{_{\perp}} ) - F_{2}(k_{_{\perp}} ) ) \nonumber \\
 & \quad + 2(m_{1} + m_{2} )(F_{3}(k_{_{\perp}} ) + F_{4}(k_{_{\perp}} ) )) + \omega_{2}((5m_{1} + m_{2} ) (F_{1}(k_{_{\perp}} ) + F_{2}(k_{_{\perp}} ) ) \nonumber \\
 & \quad + 2(e_{1} + e_{2} )(F_{3}(k_{_{\perp}} ) - F_{4}(k_{_{\perp}} ) ))\bigg](\frac{3(\vec{k} \cdot \vec{q} )^{3}}{k_{_{\perp}}^{6}} - \frac{q_{_{\perp}}^{2}\vec{k} \cdot \vec{q}}{k_{_{\perp}}^{4}} ) \bigg\};
\end{align}

\begin{align}
M F_{4}(q_{_{\perp}} ) & = - (\omega_{1} + \omega_{2} )F_{4}(q_{_{\perp}} ) - \int\frac{d\vec{k}}{(2\pi)^{3}}\frac{1}{24\omega_{1}\omega_{2}}\bigg\{ - 10{(V_S+V_V)}\bigg[\bigg(- \frac{m_{1} - m_{2}}{m_{1} + m_{2}}(e_{1} - e_{2} )\nonumber \\
 & \quad \times (F_{1}(k_{_{\perp}} ) + F_{2}(k_{_{\perp}} ) ) + \frac{e_{1} - e_{2}}{m_{1} + m_{2}}(F_{3}(k_{_{\perp}} ) - F_{4}(k_{_{\perp}} ) )\bigg)(m_{2}\omega_{1} - m_{1}\omega_{2} )\nonumber \\
 & \quad + \bigg( \frac{e_{1} - e_{2}}{m_{1} + m_{2}}(F_{1}(k_{_{\perp}} ) - F_{2}(k_{_{\perp}} ) ) + (F_{3}(k_{_{\perp}} ) + F_{4}(k_{_{\perp}} ) ) \bigg)(q_{_{\perp}}^{2} + m_{1}m_{2} - \omega_{1}\omega_{2} )\bigg]\frac{(\vec{k} \cdot \vec{q} )^{2}}{k_{_{\perp}}^{4}}\nonumber \\
 & \quad -8{(V_S-V_V)}\frac{(e_{2}m_{1} + e_{1}m_{2} )}{(e_{1} + e_{2} )(m_{1} + m_{2} )}\bigg[ m_{2}((e_{1} - e_{2} )(F_{1}(k_{_{\perp}} ) - F_{2}(k_{_{\perp}} ) ) + (m_{1} + m_{2} )\nonumber \\
 & \quad \times (F_{3}(k_{_{\perp}} ) + F_{4}(k_{_{\perp}} ) ) ) - \omega_{2}((m_{1} - m_{2} )(F_{1}(k_{_{\perp}} ) + F_{2}(k_{_{\perp}} ) ) + (e_{1} + e_{2} )(F_{3}(k_{_{\perp}} ) - F_{4}(k_{_{\perp}} ) ) )\bigg]\nonumber \\
 & \quad \times \frac{q_{_{\perp}}^{2}\vec{k} \cdot \vec{q}}{k_{_{\perp}}^{4}} +10{(V_S-V_V)}(e_{2}m_{1} + e_{1}m_{2} )\bigg[\frac{e_{1} - e_{2}}{e_{1} + e_{2}}(F_{1}(k_{_{\perp}} ) - F_{2}(k_{_{\perp}} ) ) + \frac{m_{1} + m_{2}}{e_{1} + e_{2}}\nonumber \\
 & \quad \times (F_{3}(k_{_{\perp}} ) + F_{4}(k_{_{\perp}} ) ) - \frac{\omega_{1} + \omega_{2}}{m_{1} + m_{2}}(\frac{m_{1} - m_{2}}{e_{1} + e_{2}}(F_{1}(k_{_{\perp}} ) + F_{2}(k_{_{\perp}} ) ) + (F_{3}(k_{_{\perp}} ) - F_{4}(k_{_{\perp}} ) ) )\bigg]\nonumber \\
 & \quad \times \frac{q_{_{\perp}}^{2}\vec{k} \cdot \vec{q}}{k_{_{\perp}}^{4}} -2{(V_S-V_V)}(e_{2m_{1}} + e_{1}m_{2} )\bigg[\frac{5e_{1} + e_{2}}{e_{1} + e_{2}}(F_{1}(k_{_{\perp}} ) - F_{2}(k_{_{\perp}} ) ) + 2\frac{m_{1} + m_{2}}{e_{1} + e_{2}}\nonumber \\
 & \quad \times (F_{3}(k_{_{\perp}} ) + F_{4}(k_{_{\perp}} ) ) - \frac{\omega_{1} + \omega_{2}}{m_{1} + m_{2}}(\frac{5m_{1} + m_{2}}{e_{1} + e_{2}}(F_{1}(k_{_{\perp}} ) + F_{2}(k_{_{\perp}} ) ) + 2(F_{3}(k_{_{\perp}} ) - F_{4}(k_{_{\perp}} ) ) )\bigg]\nonumber \\
 & \quad \times \frac{q_{_{\perp}}^{2}\vec{k} \cdot \vec{q}}{k_{_{\perp}}^{4}} +3{(V_S+V_V)}\bigg[- (F_{1}(k_{_{\perp}} ) + F_{2}(k_{_{\perp}} ) )(m_{1}\omega_{2} - m_{2}\omega_{1} ) - \frac{e_{1} - e_{2}}{m_{1} + m_{2}}(F_{1}(k_{_{\perp}} ) - F_{2}(k_{_{\perp}} ) ) \nonumber \\
 & \quad \times (- q_{_{\perp}}^{2} + m_{1}m_{2} - \omega_{1}\omega_{2} ) \bigg](\frac{q_{_{\perp}}^{2}}{k_{_{\perp}}^{2}} - \frac{3(\vec{k} \cdot \vec{q} )^{2}}{k_{_{\perp}}^{4}} )+ {(V_S+V_V)}\frac{1}{m_{1} + m_{2}}\bigg[-\frac{e_{1} - e_{2}}{e_{1} + e_{2}}((m_{1} - m_{2} )\nonumber \\
 & \quad \times (F_{1}(k_{_{\perp}} ) + F_{2}(k_{_{\perp}} ) ) + (e_{1} + e_{2} )(F_{3}(k_{_{\perp}} ) - F_{4}(k_{_{\perp}} ) ) ) (m_{2}\omega_{1} - m_{1}\omega_{2} ) + ((e_{1} - e_{2} )\nonumber \\
 & \quad \times (F_{1}(k_{_{\perp}} ) - F_{2}(k_{_{\perp}} ) ) + (m_{1} + m_{2} )(F_{3}(k_{_{\perp}} ) + F_{4}(k_{_{\perp}} ) ) ) (q_{_{\perp}}^{2} + m_{1}m_{2} - \omega_{1}\omega_{2} )\bigg]\nonumber \\
 & \quad \times (\frac{3q_{_{\perp}}^{2}}{k_{_{\perp}}^{2}} {+} \frac{(\vec{k} \cdot \vec{q} )^{2}}{k_{_{\perp}}^{4}} ) +2{(V_S-V_V)}\frac{e_{2}m_{1} + e_{1}m_{2}}{(e_{1} + e_{2} )(m_{1} + m_{2} )}\bigg[ m_{2}((5e_{1} + e_{2} ) (F_{1}(k_{_{\perp}} ) - F_{2}(k_{_{\perp}} ) ) \nonumber \\
 & \quad + 2(m_{1} + m_{2} )(F_{3}(k_{_{\perp}} ) + F_{4}(k_{_{\perp}} ) )) - \omega_{2}((5m_{1} + m_{2} ) (F_{1}(k_{_{\perp}} ) + F_{2}(k_{_{\perp}} ) ) \nonumber \\
 & \quad + 2(e_{1} + e_{2} )(F_{3}(k_{_{\perp}} ) - F_{4}(k_{_{\perp}} ) ))\bigg](\frac{3(\vec{k} \cdot \vec{q} )^{3}}{k_{_{\perp}}^{6}} - \frac{q_{_{\perp}}^{2}\vec{k} \cdot \vec{q}}{k_{_{\perp}}^{4}} )\bigg\}.
\end{align}
Here, $e_i=\sqrt{m_i^2+k_{T}^2}$, {$V_S=V_S(\vec{q}-\vec{k})$, and  $V_V=V_V(\vec{q}-\vec{k})$}. {When solving the Salpeter equations Eqs.(A1-A4), since the radial wave function $\zeta_i(\vec{q}^2)$ or $F_i(\vec{q}^2)$ decreases with the increase of $|\vec{q}|$ (see Fig.3 for example), we truncate the relative momentum $|\vec{q}|$ (and $|\vec{k}|$) to a certain maximum value $|\vec{q}|_{max}$ ($|\vec{k}|_{max}=|\vec{q}|_{max}$), and discretize this momentum into $n$ parts ($n$ is a large number). Thus the four coupled eigenvalue equations were transformed into a $4n \times 4n$ matrix formula, and then the numerical solutions were implemented.}

\end{document}